\documentclass[a4paper,fleqn,usenatbib]{mn2e}

\usepackage{amsmath}	
\usepackage{amssymb}	

\usepackage{aas_macros}

\usepackage[T1]{fontenc}

\usepackage{graphicx}	
\usepackage{longtable,lscape,afterpage}

\newcommand{\teff}{$T_{\rm eff}$}

\newcommand{\ha}{H$\alpha$}
\newcommand{\hb}{H$\beta$}
\newcommand{\vs}{$V_{\rm rot}\sin i$}

\newcommand{\lgg}{$\log~g$}
\newcommand{\ion}[2]{#1~\textsc{#2}}

\newcommand{\Vmic}{$V_{\rm mic}$}
\newcommand{\Vmac}{$V_{\rm mac}$}
\newcommand{\kH}{$S_{\rm H}$}

\newcommand{\header}[1]{\multicolumn{1}{|c|}{\textbf{#1}}}
\newcommand{\headerb}[1]{\multicolumn{1}{|l|}{\textbf{#1}}}

\title[Parameters determination in FGK dwarfs]
{Accuracy of atmospheric parameters of FGK dwarfs determined by spectrum fitting
}

\author[T.~Ryabchikova et al.]
{
\parbox{\textwidth}{
T.~Ryabchikova$^1$\thanks{E-mail: ryabchik@inasan.ru}, 
N.~Piskunov$^{2}$, Yu.~Pakhomov$^{1}$, V.~Tsymbal$^{3}$, A.~Titarenko$^{1,4}$,
T.~Sitnova$^{1,4}$, S.~Alexeeva$^{1}$, L.~Fossati$^{5}$, L.~Mashonkina$^{1}$
} \\
\\ 
$^{1}$Institute of Astronomy, Russian Academy of Sciences, Pyatnitskaya 48,
119017 Moscow, Russia\\
$^{2}$Department of  Physics and Astronomy, Division of Astronomy and Space
Physics, Uppsala University, Box 516, 751 20 Uppsala, Sweden\\
$^{3}$Physical Technical Institute, Crimean Federal University, Vernadskiy's
Avenue 4, 95007 Simferopol, Crimea\\
$^{4}$Department of Astronomy, Physics Faculty, M.V.Lomonosov Moscow State
University, GSP-1, 1-2 Leninskye Gory, 119991 Moscow, Russia\\
$^{5}$Space Research Institute, Austrian Academy of Sciences, Schmiedlstrasse 	
6, A-8042 Graz, Austria
}

\date{Accepted XXX. Received YYY; in original form ZZZ}

\pubyear{2015}

\begin{document}
\label{firstpage}
\pagerange{\pageref{firstpage}--\pageref{lastpage}}
\maketitle

\begin{abstract}
We performed extensive tests of the accuracy of atmospheric parameter
determination for FGK stars based on the spectrum fitting procedure Spectroscopy
Made Easy (SME). Our stellar sample consists of 13 objects, including the Sun, in
the temperature range 5000--6600~K and metallicity range $-$1.4 -- $+$0.4. The
analysed stars have the advantage of having parameters derived by
interferometry. For each star we use spectra obtained with different
spectrographs and different signal-to-noise
ratios (S/N). For the fitting we adopted three different sets of constraints and
test how the derived parameters depend upon the spectral regions
(masks) used in SME. We developed and implemented in SME a new method for
estimating uncertainties in the resulting parameters based on fitting residuals,
partial derivatives, and data uncertainties. For stars in the 5700--6600 K range
the best agreement with the effective temperatures derived by interferometry is
achieved when spectrum fitting includes the H$\alpha$ and H$\beta$ lines, while
for cooler stars the choice of the mask does not affect the results. The derived
atmospheric parameters do not strongly depend on spectral
resolution and S/N of the observations, while the uncertainties in temperature
and surface gravity increase with increasing effective temperature, with minima
at 50~K in \teff\ and 0.1~dex in \lgg, for spectra with S/N=150--200. A NLTE
analysis of the \ion{Ti}{i}/\ion{Ti}{ii} and \ion{Fe}{i}/\ion{Fe}{ii} 
ionisation equilibria
and abundances determined from the atomic \ion{C}{i} (NLTE) and molecular CH
species supports the parameters we derived with SME by fitting the observed
spectra including the hydrogen lines.     
\end{abstract}




\section{Introduction}
\label{intro}
The discovery of exoplanets, the results of the Kepler space mission, and the
expected huge amount of information from the Gaia mission have attracted special
attention to the problem of accurately determining atmospheric parameters for a
large number of stars. Accurate atmospheric parameters are essential for
exoplanet studies to best derive the planet parameters (e.g., mass, radius, and
equilibrium temperature) and to characterise their atmospheres. The Gaia
mission\footnote{http://sci.esa.int/gaia/} will provide unprecedented positional
and proper motion measurements for about one billion stars in our Galaxy and
throughout the Local Group and, thus, accurate parallaxes needed to improve
stellar surface gravities. Accurate kinematic properties together with accurate
atmospheric parameters will allow one to constrain Galactic stellar populations
and chemical evolution more accurately than ever been possible before. This
space mission is supported by ground-based spectroscopic observations within the
Gaia-ESO Survey (GES) project \citep{2012Msngr.147...25G}. Determination of
atmospheric parameters for the GES benchmark and program stars was discussed by
\citet{2014A&A...570A.122S} and \citet{2015arXiv150606095H}.

Determining atmospheric parameters for a large number of stars requires the
development of automatic procedures capable of quickly producing reliable
results. Following this reasoning, several different automatic spectral analysis
procedures were developed during the last decades. All of them are essentially
based on either the fit of synthetic to observed spectra or the measurement and
analysis of equivalent widths ($EW$) of metal lines. For F-, G-, and K-type
stars and for both techniques various authors claimed a very high accuracy of
the order of 20--40~K in effective temperature, 0.02--0.06~dex in surface
gravity, and 0.02--0.05~dex in metallicity derived in the spectral analysis
with S/N$\sim$150-400 \citep{2005ApJS..159..141V, 2006AA...458..873S}.
Nevertheless, it has been shown that analysis of the spectra obtained using the
same spectrograph, but in different observational runs may lead to results that
differ in effective temperature by 70~K, though the spectra had been analysed
using the same technique \citep[e.g., see Table~3 of][]{2010MNRAS.403.1368G}.
The situation is getting worse, when analysing the same spectra with different
techniques.  \citet{2012ApJ...757..161T} compared atmospheric parameters derived
by the spectrum fitting and the $EW$ techniques, and for stars hotter than
6000~K they found that the values from the spectrum fitting method are
systematically lower compared to those from the $EW$ method.
\citet{2014A&A...570A.122S} showed the presence of a large spread in the
atmospheric parameters derived by thirteen different groups using different
techniques for 11 bright F-, G-, and K-type main-sequence stars (the Gaia
FGK benchmark stars). The Gaia spectral library collects spectra for these stars
\citep{2014A&A...566A..98B}. Most of them have S/N$>$200. The average
difference between the recommended effective temperatures and gravities and
individual determinations from the six different spectrum fitting and seven $EW$
methods ranges between  46~K and 186~K and 0.09~dex and 0.28~dex, respectively.
The largest discrepancies are probably caused by a systematic
component, which is usually difficult to accurately estimate. It is worth
noting, the difference in parameters reaches 100~K in effective temperature and
0.16~dex in surface gravity even between the two groups using a common technique
based on $EW$. The analysis of the results for all groups of FGK dwarfs led to a
method-to-method dispersion of $\pm$150~K in \teff, $\pm$0.30~dex in \lgg, and
$\pm$0.10~dex in [Fe/H]. The situation is even worse for giant and metal-poor
stars.  These examples show how far the actual uncertainties may differ from the
adopted ones.

This paper presents an extensive analysis of the uncertainties in atmospheric
parameter determination made with the fitting procedure Spectroscopy Made Easy
\citep[SME,][]{1996AAS..118..595V}. The investigated stars and their spectra are
described in Section~\ref{obs}. A short description of the SME tool, the
implemented modifications, and the error analysis are given in
Section~\ref{sme}. We tested the best recommended atmospheric parameters and
their uncertainties in Section~\ref{spectr} by inspecting the
\ion{Ti}{i}/\ion{Ti}{ii} and \ion{Fe}{i}/\ion{Fe}{ii} ionisation equilibrium and
element abundances from the atomic \ion{C}{i} and molecular CH lines.
Section~\ref{concl} summarises our conclusions.

\section{Observations}
\label{obs}

For the spectroscopic analysis we choose the 13 main-sequence (MS) stars 
including the Sun (Table~\ref{table1}) in the 4900--6600~K temperature range
and with metallicity between [Fe/H] = $-$1.5 and $+$0.3~dex. All the stars,
except HD~149026, have, at least, one interferometric determination of radius
and effective temperature. Our methods of atmospheric parameter
determination were first applied to the Sun that is the only star with a
directly measured effective temperature and surface gravity.

Spectra of the program stars were obtained with different spectrographs.
Most data were extracted from the following archives: the UVES/VLT and HARPS/3.6
m spectrographs at
ESO\footnote{http://archive.eso.org/eso/eso\_archive\_main.html}, the
ELODIE/1.93-m spectrograph\footnote{http://atlas.obs-hp.fr/elodie/} at the
Observatoire de Haute Provence, and the ESPaDONs spectrograph at the
Canada-France-Hawaii
Telescope\footnote{http://www.cadc-ccda.hia-iha.nrc-cnrc.gc.ca/en/cfht/} (CFHT).
Spectra of $\beta$~Vir and HD~103095 were obtained with the FOCES spectrograph
at 2.2-m telescope of the Calar Alto Observatory \citep{1998AA...338..161F}. One
of the spectra of 61~Vir was obtained with the Hamilton Echelle Spectrograph
attached to the Shane 3-m telescope of the Lick Observatory
\citep{2015ApJ...808..148S}. Spectra of few stars, including that of the Sun
reflected from Ganymede, were obtained with the HiReS/Keck spectrograph
\citep{2010ApJ...721.1467H}. All the retrieved spectra have been reduced with
the standard reduction pipelines available for each spectrograph. For the Sun,
we used the National Solar Observatory (NSO) solar flux spectrum
\citep{1984sfat.book.....K}.

For each star we tried to find an archival spectrum with as high as
possible signal-to-noise ratio (S/N). All observed spectra have a resolving
power R$>$40\,000 with a peak resolution of R$\sim$110\,000 obtained with the
UVES and HARPS spectrographs. Most ESPaDONs spectra have been collected in
spectropolarimetric mode, which provides spectra with a resolution of R =
65\,000 instead of R = 80\,000 reached with the ``object only'' mode. Rather
large range of S/N and R for the collected spectra allows us to thoroughly
explore the uncertainties in the atmospheric parameters as a function of the
basic spectral characteristics. Detailed information on the spectral resolution
and S/N along with the program ID is given in Table~\ref{table1} for each
analysed spectrum. Continuum rectification was performed using low-order
polynomials, giving special attention to the regions covered by the H$\beta$ and
H$\alpha$ lines, which play a crucial role in the atmospheric parameter
determination. 

\afterpage{
\onecolumn
\begin{landscape}
\LTcapwidth=\textheight
\LTleft=0cm
\setlength{\tabcolsep}{0.5mm}
\renewcommand\arraystretch{1.1}
\begin{longtable}{lrllrllclllrl}
\caption{List of investigated stars and spectrographs used to collect the
spectra. The first column lists also the resolving power of each spectrum.
Column 2 gives the average S/N calculated in the three wavelength ranges
discussed in Sect.~\ref{masks}. The atmospheric parameters derived with SME are
given in columns 4, 5, and 6, while columns 7 and 8 list the atmospheric
parameters derived from interferometry. The last four columns indicate the
spectroscopic atmospheric parameters from the literature and their source. The
uncertainties are given in parentheses.}
\label{table1}\\

\hline
\multicolumn{2}{|c}{Observations} & 
\multicolumn{4}{|c}{SME} &
\multicolumn{3}{|c}{Interferometry} & 
\multicolumn{4}{|c|}{Other spectroscopy}   \\
\hline
\multicolumn{1}{|c|}{Spectrograph} &  \header{S/N}  & \header{mask} & 
\header{\teff,K} & \header{\lgg} & \header{[M/H]} & 
\header{\teff,K} & \header{\lgg} & \header{Reference} & 
\header{\teff,K} & \header{\lgg} & \header{[M/H]} & \header{Reference}  \\
\multicolumn{1}{|c|}{Resol.
power}&\header{}&\header{}&\header{}&\header{}&\header{}&\header{}&\header{}
&\header{}&\header{}&\header{}&\header{}&\header{} \\
\multicolumn{1}{|c|}{Program
ID}&\header{}&\header{}&\header{}&\header{}&\header{}&\header{}&\header{}
&\header{}&\header{}&\header{}&\header{}&\header{} \\
\hline
 \multicolumn{1}{c}{(1)} & \multicolumn{1}{c}{(2)} & \multicolumn{1}{c}{(3)} &
\multicolumn{1}{c}{(4)} & \multicolumn{1}{c}{(5)} & \multicolumn{1}{c}{(6)} &
\multicolumn{1}{c}{(7)} & \multicolumn{1}{c}{(8)} & \multicolumn{1}{c}{(9)} &
\multicolumn{1}{c}{(10)} & \multicolumn{1}{c}{(11)} & \multicolumn{1}{c}{(12)} &
\multicolumn{1}{c}{(13)}\\ 
\endfirsthead

\caption{continue.}\\
\hline
 \multicolumn{1}{c}{(1)} & \multicolumn{1}{c}{(2)} & \multicolumn{1}{c}{(3)} &
\multicolumn{1}{c}{(4)} & \multicolumn{1}{c}{(5)} & \multicolumn{1}{c}{(6)} &
\multicolumn{1}{c}{(7)} & \multicolumn{1}{c}{(8)} & \multicolumn{1}{c}{(9)} &
\multicolumn{1}{c}{(10)} & \multicolumn{1}{c}{(11)} & \multicolumn{1}{c}{(12)} &
\multicolumn{1}{c}{(13)}\\ 
\endhead

\hline
\multicolumn{13}{c}{{\bf HD~61421 = Procyon}} \\
\hline
 UVES            & 500   & m6   & 6615(89)  & 3.89(33) &  -0.05(05) &  6597(18) 
& 4.00(02)  & \citet{2013ApJ...771...40B} &6485(80)  & 3.89(09) &  0.01(07)& 
\citet{2010MNRAS.405.1907B}       \\
 (80000)         &	    & m5   & 6579(93)  & 3.84(37) &  -0.07(05) & 
6573(42)  & 4.00(02)  & \citet{2013ApJ...771...40B} &6593(50)  & 3.90(01) & 
0.02(03)&  \citet{2010MNRAS.403.1368G}     \\
UVES POP         &	    & m4   & 6690(89)  & 3.85(25) &  -0.02(05) & 
6563(33)  & 4.00(02)  & \citet{2013ApJ...771...40B} &6660(95)  & 4.05(06) & 
0.02(09)&  \citet{2013MNRAS.428.3164D}        \\
\citep{2003Msngr.114...10B}&&VF & 6602(112) & 3.98(17) &  -0.11(07) &  6562(32) 
& 4.00(02)  & \citet{2013ApJ...771...40B} &          &          &          &    
   \\
\hline
\multicolumn{13}{c}{{\bf HD~49933}} \\
\hline
 HARPS           & 350   & m6   & 6582(115) & 4.00(52) &  -0.48(09) &  6635(90) 
& 4.21(05)  & \citet{2013ApJ...771...40B} &6570(60)  & 4.28(06) & -0.44(03)& 
\citet{2009AA...506..235B}   	\\ 
 (110000)        &	    & m4   & 6653(143) & 4.08(33) &  -0.44(08) &	
    &	         &  &6600(80)  & 4.15(05) & -0.47(07)& 
\citet{2015ApJ...808..148S}  		\\
076.C-0279       &	    & VF   & 6512(120) & 4.13(20) &  -0.57(10) &	
    &	         &  &		&	    &	        & 		 	
                     \\
\cline{1-6}
 ESPaDONs        & 300   & m6   & 6546(100) & 4.04(52) &  -0.50(08) &		
   &	         &  &		&	    &	        & 		  	
                    \\
 (65000)         &	    & m4   & 6706(149) & 4.22(33) &  -0.41(08) &	
    &	         &  &		&	    &	        & 		  	
                    \\	   
05bf6a           &	    & VF   & 6567(107) & 4.17(15) &  -0.54(08) &	
    &		  &  &		&	    &	        & 		  	
                    \\    
\hline
\multicolumn{13}{c}{{\bf HD~9826 = $\upsilon$~And}} \\
\hline
 ESPaDONs        & 700   & m6   & 6145(39)  & 4.06(12) &   0.05(02) &  6177(25) 
& 4.13(03)  & \citet{2013ApJ...771...40B} &6170(48)  & 4.00(08) &  0.08(04)& 
\citet{2010MNRAS.403.1368G}  		 \\
 (65000)         &	    & m4   & 6323(38)  & 4.25(12) &   0.15(03) & 
6027(26)  & 4.10(02)  & \citet{2013ApJ...771...40B} &6239(37)  & 4.19(03) & 
0.14(03)&  \citet{2010MNRAS.403.1368G}  		 \\
05BO2            &	    & VF   & 6292(27)  & 4.22(04) &   0.07(03) &	
    &	         &  & 	       &	    &	        &		   	
                      \\
\cline{1-6}
 ELODIE          & 250   & m6   & 6132(84)  & 4.03(30) &   0.14(06) &		
   &	         &  & 	       &	    &	        &		  	
                       \\
(42000)          &       &      &           &          &            &           
&           &  &          &          &          &                             \\
 \cline{1-6}
 HiReS	          & 200   & m4   & 6354(103) & 4.33(28) &   0.13(07) &		
   &	         &  & 	       &	    &	        &		   	
                      \\
 (69000)         &	    & VF   & 6278(76)  & 4.20(09) &   0.08(07) &	
    &	         &  & 	       &	    &	        &		   	
                      \\
\hline
\multicolumn{13}{c}{{\bf HD~102870 = $\beta$~Vir}} \\
\hline
 ESPaDONs        & 1000  & m6   & 6122(27)  & 4.07(10) &   0.10(02) &  6054(13) 
& 4.11(04)  & \citet{2013ApJ...771...40B} &6050(80)  & 3.98(07) &  0.12(07)& 
\citet{2010MNRAS.405.1907B}   	\\ 
  (65000)        &	    & m4   & 6232(27)  & 4.18(09) &   0.18(02) &        
   &           &  &6111(28)  & 4.00(05) &  0.16(02)& 
\citet{2010MNRAS.403.1368G}  		\\
05bf6a           &	    & VF   & 6214(21)  & 4.13(03) &   0.12(02) &	
    &	         &  &6180(36)  & 4.15(05) &  0.21(03)& 
\citet{2010MNRAS.403.1368G}  		\\
 \cline{1-6}
 FOCES	          & 250   & m6   & 6122(85)  & 4.02(30) &   0.13(11) &		
   &	         &  &6170(80)  & 4.14(04) &  0.11(06)& 
\citet{2015ApJ...808..148S}   		\\
 (60000)         &	    & m4   & 6242(81)  & 4.14(23) &   0.21(06) &	
    &	         &  &		&	    &	      	 &		        
                 \\
             	   &	    & VF   & 6218(75)  & 4.10(13) &   0.14(04) &	
    &	         &  &		&	    &	        &		        
                 \\
\hline\pagebreak
\hline
\multicolumn{13}{c}{{\bf HD~149026}}\\*
\hline
 ESPaDONs        & 200   & m6   & 6074(82)  & 4.18(29) &   0.24(06) &	        
  &	         &  &6131(35)  & 4.22(05) &  0.31(03)& 
\citet{2010MNRAS.403.1368G} \\*
 (65000)         &	    & m4   & 6239(74)  & 4.33(23) &   0.33(06) &	
          &	         &  &6103(66)  & 4.27(05) &  0.24(07)& 
\citet{2012ApJ...757..161T} \\*
07ah28a          &	    & VF   & 6183(62)  & 4.22(09) &   0.27(05) &	
    &	         &  &		&	    &	        & \\*
 \cline{1-6}
 HiReS	          & 300   & m4   & 6285(95)  & 4.46(31) &   0.37(07) &		
   &	         &  &		&	    &	        & \\*
 (69000)         &	    & VF   & 6137(85)  & 4.23(17) &   0.28(07) &	
    &	         &  &		&	    &	        &  \\*
\hline
\multicolumn{13}{c}{{\bf HD~209458}}\\
\hline
 ESPaDONs        & 1250  & m6   & 6033(28)  & 4.28(08) &  -0.05(02) &  6092(103)
&  4.28(10) & \citet{2015MNRAS.447..846B} &6118(25)  & 4.50(04) &  0.03(02)& 
\citet{2008AA...487..373S}	  		\\
 (80000)         &	    & m4   & 6160(27)  & 4.39(07) &   0.02(07) &	
    &	         &  &6117(26)  & 4.48(08) &  0.02(03)& 
\citet{2004AA...415.1153S}     		\\
co-added\footnote{05BC16, 05BH32A, 06AC12, 06AH20A}  
                 &	    & VF   & 6145(15)  & 4.41(02) &  -0.05(02) &	
    &	         &  &          &          &          &                          
   \\
    &	         &  &6065(50)  & 4.42(04) &  0.00(05)& 
\citet{2008ApJ...677.1324T} \\
\cline{1-6}
 HARPS           & 300   & m6   & 6010(74)  & 4.23(22) &  -0.08(05) &		
   &	         &  &		&	    &	        & 		  	
\\
 (110000)        &	    & m4   & 6116(65)  & 4.35(16) &  -0.01(04) &	
    &	         &  &		&	    &	        & 		  	
\\
co-added\footnote{076.C-0878, 183.C-0972, 074.C-0012, 60.A-9036}
                 &       & VF   & 6112(45)  & 4.38(06) &  -0.07(04) &		
   &	         &  &		&	    &	        & 		  	
\\
 \cline{1-6}
 UVES	          & 300   & m6   & 5987(70)  & 4.25(23) &  -0.09(05) &		
   &	         &  &	       &	    & 		 &		\\
 (110000)        &	    & m4   & 6106(67)  & 4.38(17) &  -0.02(05) &	
    &	    	  &  &	       &	    & 		 &		\\
co-added\footnote{265.C-5038, 067.C-0206, 077.C-0379, 087.D-0010, 088.C-0879}
                 &	    & VF   & 6093(42)  & 4.38(06) &  -0.07(04) &	
    &	    	  &  &	       &	    & 		 &		\\
 \cline{1-6}
 HiReS	          & 170   & m4   & 6174(140) & 4.43(39) &   0.02(10) &		
   &	    	  &  &	       &	    & 		 &		\\
 (69000)         &	    & VF   & 6181(103) & 4.42(18) &  -0.03(08) &	
    &	         &  &		&	    & 		 &		\\
 \cline{1-6}
 ELODIE          & 170   & m6   & 5980(123) & 4.19(40) &  -0.08(09) &		
   &	         &  &		&	    & 		 &		\\
 (42000)         &	    & m4   & 6152(120) & 4.36(33) &   0.02(09) &	
    &	         &  &		&	    & 		 &		\\
\hline
\multicolumn{13}{c}{{\bf Sun}} \\
\hline
 Atlas          & 1000 & m6	  & 5757(24)  & 4.41(06) &  -0.03(02) &  5777	
 & 4.44      &  &  	     &	         &	         &\\
 (550000)       &	  & m4   & 5787(22)  & 4.44(05) &  -0.01(02) &		
 &	       &  &  	     &	         &	         &\\
                &	  & VF   & 5773(12)  & 4.41(02) &  -0.06(02) &		
 &	       &  &  	     &	         &	         &\\
 \cline{1-6}
 HiRes       	  & 200  & m4	  & 5778(67)  & 4.43(18) &  -0.00(06) &         
  &	       &  &  	     &	         &	         &\\
 (69000)     	  &	  & VF   & 5792(42)  & 4.46(06) &  -0.06(04) &		
 &	       &  &  	     &	         &	         &\\
\hline
\multicolumn{13}{c}{{\bf HD~1461}} \\
\hline
 UVES	         &  270 & m6   & 5732(55)  & 4.31(17) &   0.14(05) &  5386(60) 
& 4.23(05)  & \citet{2014MNRAS.438.2413V} & 5765(18)  & 4.38(03) &  0.19(01)   &
\citet{2008AA...487..373S}     		  \\
(115000)        &	  & m4   & 5764(51)  & 4.35(16) &   0.16(04) &  	
        &	       &  & 5751(50)  &	4.33(07) &  0.18(04)   &
\citet{2009AA...508L..17R}\\		  
076.B-0055      &	  & VF   & 5798(38)  & 4.36(06) &   0.13(04) &  	
        &	       &  & 5765(44)  &	4.41(06) &  0.18(03)   &
\citet{2005ApJS..159..141V}\\		  
 \cline{1-6}
 HiRes       	  &  150 & m4   & 5805(80)  & 4.41(23) &   0.18(07) &  	        
&	       &  & 	     &	         &		  &\\
 (69000)        &	  & VF   & 5837(53)  & 4.40(09) &   0.16(06) &  	
        &	       &  & 	     &	         &  		  &\\
 \cline{1-6}
 ELODIE         &  150 & m6   & 5757(92)  & 4.31(28) &   0.12(08) &  	        
&	       &  & 	     &	         &  	         &\\
 (42000)        &      & m4   & 5815(89)  & 4.36(25) &   0.17(08) &  	        
&	       &  & 	     &	         &  	         &\\
\hline
\multicolumn{13}{c}{{\bf HD~115617 = 61~Vir}} \\
\hline
 ESPaDONs       & 700  & m6   & 5567(27)  & 4.42(07) &  -0.03(02) &  5538(13)  &
4.42(03)  & \citet{2014MNRAS.438.2413V} & 5558(19)  & 4.36(03) & -0.02(01)   & 
\citet{2008AA...487..373S}	 		\\ 
 (65000)        &	  & m4   & 5595(23)  & 4.45(06) &  -0.01(02) &		
 &	       &  & 5490(80)  & 4.40(12) & -0.10(05)   & 
\citet{2015ApJ...808..148S}   		\\
11AC04          &	  & VF   & 5588(12)  & 4.46(02) &  -0.06(02) &		
 &	       &  & 	     &	         &	         &\\
\cline{1-6}
 Lick	         & 200  & m4   & 5559(69)  & 4.42(18) &  -0.02(06) &		
 &	       &  & 	     &	         &	         &\\
 (60000)        &	  & VF   & 5617(54)  & 4.45(09) &   0.02(07) &		
 &	       &  & 	     &	         &	         &\\
\hline
\multicolumn{13}{c}{{\bf HD~69830}} \\
\hline
 ESPaDONs       & 650  & m6   & 5422(43)  & 4.47(12) &  -0.04(04) &  5394(62)  &
4.46(04)  & \citet{2015ApJ...800..115T} & 5402(28)  & 4.40(04) & -0.06(02)   & 
\citet{2008AA...487..373S}	  		\\
 (65000)        &	  & m4   & 5436(39)  & 4.49(10) &  -0.03(04) &		
 &	       &  & 5413(28)  & 4.52(07) & -0.01(02)   & 
\citet{2010MNRAS.403.1368G}  		\\
14AF14          &	  & VF   & 5419(23)  & 4.48(05) &  -0.09(04) &		
 &	       &  & 5382(25)  & 4.34(09) & -0.02(02)   & 
\citet{2010MNRAS.403.1368G}    		\\
\cline{1-6}
 HiRes	         & 200  & m4   & 5424(60)  & 4.58(17) &   0.04(07) &		
 &	       &  & 5385(20)  &	4.37(02) & -0.05(02)   & 
\citet{2006AA...458..873S}		  		\\
 (69000)        &   	  & VF   & 5415(45)  & 4.49(10) &  -0.09(07) &	        
&	       &  & 	     &	         &	         &\\
\hline
\multicolumn{13}{c}{{\bf HD~189733}} \\
\hline

 ESPaDONs       & 1500 & m6   & 5049(10)  & 4.53(03) &  -0.02(01) &  4875(43)  &
4.56(03) & \citet{2015MNRAS.447..846B}  & 5051(47)  & 4.53(08) & -0.03(05)   & 
\citet{2006AA...458..873S}    		\\
 (65000)        &	  & m4   & 5076( 8)  & 4.58(03) &  -0.00(01) &		
 &	      &   & 5111(77)  & 4.59(01) & -0.04(07)   & 
\citet{2012ApJ...757..161T}    		\\
co-added\footnote{06AF34, 06bd01, 06bf27, 07AC27}
                &	  & VF   & 5016( 6)  & 4.64(02) &  -0.05(01) &		
 &	      &   & 5034(90)  & 4.54(10) & -0.03(07)   & 
\citet{2008MNRAS.384..173F}    		\\
\cline{1-6}                                                                     
                                                  
 HiRes       	  & 180  & m4   & 5056(68)  & 4.53(18) &  -0.02(10) &		
 &	      &   & 4952(64)  & 4.26(12) &	 0.01(04)   & 
\citet{2010MNRAS.403.1368G} 	         \\
 (69000)        &	  & VF   & 5019(62)  & 4.57(16) &  -0.06(09) &		
 &	      &   & 		 &	     &	            & 		  	
\\

\hline
\multicolumn{13}{c}{{\bf HD~23249 = $\delta$~Eri}} \\
\hline
 HARPS          & 500  & m5	  & 5040(28)  & 3.73(07) &   0.05(03) & 
4955(30)  & 3.77(03)  & \citet{2013ApJ...771...40B} & 5150(51)  & 3.89(08) & 
0.13(04)   &  \citet{2008AA...487..373S}      \\
 (110000)       &      & m4	  & 5037(21)  & 3.73(06) &   0.05(02) &         
  &           &  & 5015(80)  & 3.77(05) &  0.15(07)   & 
\citet{2010MNRAS.405.1907B}     \\
60.A-9036       &	  & VF   & 5053(15)  & 3.87(05) &   0.05(03) &		
 &	       &  &           &          &             &\\
\cline{1-6}
 HiRes	         & 200  & m4	  & 5052(48)  & 3.76(14) &   0.07(06) &		
 & 	       &  &  	     &	         &	         &\\
 (69000)        &	  & VF   & 5085(27)  & 3.78(07) &   0.05(05) &		
 &	       &  &  	     &	         &	         &\\
\hline
\multicolumn{13}{c}{{\bf HD~103095}} \\
\hline
 FOCES       	  & 150  & m6	  & 4958(46)  & 4.52(17) &  -1.38(10) & 
4771(18)  & 4.56(03)  & \citet{2013ApJ...771...40B} & 5130(65)  & 4.66(06) &
-1.26(08)   &\citet{2015ApJ...808..148S}   	 \\
(60000)         &	  & m4   & 4944(43)  & 4.48(15) &  -1.36(09) &  4831(25)
 & 4.58(04)  & \citet{2013ApJ...771...40B} & 5110(80)  & 4.66(10) & -1.35(05)  
&\citet{1998AA...338..161F}\\
             	  &	  & VF   & 4904(27)  & 4.49(07) &  -1.52(06) &		
 &	       &  & 4930(44)  & 4.65(06) & -1.37(03)  
&\citet{2005ApJS..159..141V}\\
\hline
\end{longtable}
\end{landscape}
\twocolumn
}

\section{Atmospheric parameter determination}
\label{sme}

We determined the stellar atmospheric parameters with the SME package
\citep{1996AAS..118..595V}. The tool SME was designed to perform an analysis of
stellar spectra using spectrum fitting techniques in a consistent and
reproducible way. The fit can be done in several spectral intervals
simultaneously. The data points can further be masked to avoid observational
defects and/or spectral regions with uncertain atomic/molecular data. Since
the first version of SME,
a number of modifications and improvements were made to the original version.
They include new molecular and negative hydrogen ion partition functions and
equilibrium constants in the equation-of-state (EOS) package, a new algorithm
for solving the radiative transfer \citep[the Feautrier algorithm was replaced
with a Bezier attenuation operator scheme; see][]{2013ApJ...764...33D}, and a
modification of the interpolation procedure in the model atmosphere grid.
Further details of the current version of SME can be found in the upcoming paper
by \citet{SME2015ApJ}. In this work, we use the sme\_443 package version. The
SME package is working with various grids of model atmospheres: ATLAS9 and
ATLAS12 \citep{2004astro.ph..5087C}, MARCS \citep{2008AA...486..951G} for
dwarfs, and LLmodels \citep{2004AA...428..993S}. All these models assume
plane-parallel one-dimensional (1D) geometry and the local thermodynamical
equilibrium (LTE). Our calculations were performed with the MARCS models.

\subsection{Choice of spectral windows}
\label{masks}

For stars hotter than 6000~K \citet{2012ApJ...757..161T} found that the SME and
SPC \citep[Stellar Parameter Classification,][]{2012Natur.486..375B} spectrum
fitting procedures produce systematically lower effective temperatures and
surface gravities compared to the results based on the $EW$ measurements and
analysis of the \ion{Fe}{i} excitation and \ion{Fe}{i}/\ion{Fe}{ii} ionisation
equilibria. These authors used spectral intervals similar to those adopted by
\citet{2005ApJS..159..141V}, which include the 5163--5190~\AA\ spectral region
and few small windows in the 6000--6180~\AA\ wavelength range. The first region
contains the \ion{Mg}{i}b lines, whose extended wings are sensitive to surface
gravity variations in G- and K-type stars, while being less sensitive for F-type
stars. Most of the other lines belong to Fe-peak elements. The strong
\ion{Ca}{i} lines in the 6100--6180~\AA\ spectral range were not included in the
fitting. 

 For the present analysis we have selected four spectral regions to be
fitted by SME: 4485--4590~\AA, 5100--5200~\AA, 5600--5700~\AA, and
6100--6200~\AA. They include the spectral features, which are sensitive to a
variation in different atmospheric parameters . The 4485--4590~\AA\ spectral
range includes numerous \ion{Ti}{i}-\ion{Ti}{ii} and \ion{Fe}{i}-\ion{Fe}{ii}
lines with accurate laboratory data, making this region very sensitive to
gravity variations. The 5100--5200~\AA\ spectral range covers the molecular
C$_2$ (C$_2$ Swan system) and MgH lines, which are strongly temperature
dependent in cool stars. The 5167--5183~\AA\ and 6100-6200~\AA\ intervals
include the \ion{Mg}{i}b and strong \ion{Ca}{i} lines with Lorentz wings induced
by the collisional broadening sensitive to gravity variations.

However, in spectra of stars hotter than the Sun the line broadening due to
collisions with hydrogen atoms (i.e., Van der Waals broadening) becomes smaller
and the molecular lines disappear. Therefore, the degeneracy between different
atmospheric parameters increases, resulting in biases in temperature, gravity,
and metallicity. To solve this problem, one needs to use features that are
strongly sensitive to one of the parameters, for example the hydrogen Balmer
line wings, which are temperature indicators in F-, G-, and K-type stars
\citep{1993A&A...271..451F, 2011A&A...531A..83C}. When dealing with hydrogen
lines, the major problem is a correct continuum normalisation, in particular for
high-resolution echelle spectra. We included the H$\beta$ (4820--4880~\AA) and
H$\alpha$ (6520--6580~\AA) lines in the SME fitting, excluding the $\pm$1.5~\AA\
regions around the line centers. Also the cores of the strong \ion{Mg}{i} and
\ion{Ca}{i} lines were not considered in the fit because of departures from LTE.

The SME mask was used to remove unidentified features and spectral lines with
uncertain atomic parameters. The mask was verified using the solar atlas and
then applied to all stars. For individual observations the mask was further
corrected to exclude regions showing spectral defects (e.g., dead pixels),
remaining cosmics, and telluric lines, particularly in the wings of the
H$\alpha$ line. The use of new/extended spectral regions was made possible
thanks to a new version of the Vienna Atomic Line
Database\footnote{http://vald.astro.univie.ac.at/~vald3/php/vald.php}
\citep[VALD3;][]{2015PhyS...90e4005R} that includes a new extensive set of
atomic line calculations by R.~Kurucz
\footnote{http://kurucz.harvard.edu/atoms.html} and his collection of molecular
lines. Also the new data for the C$_2$ Swan system from \citet{BBSB} were
included. For lines of Sc, V, Mn, Co, and Cu the full hyperfine structure (HFS)
was taken into account (see references on Kurucz's web site). Hydrogen line
profiles were calculated with the code described in \citet[][and references
therein]{2000A&A...363.1091B}\footnote{
http://www.astro.uu.se/~barklem/hlinop.html}. For comparison purposes we also
run SME using the mask adopted by \citet{2005ApJS..159..141V}, including their
atomic and molecular line parameters. 

We analysed the stars using the four following masks:

\begin{enumerate}
\item the four regions covering the metal lines (hereafter, m4), 
\item m4 plus the H$\alpha$ line (hereafter, m5), 
\item m4 plus the H$\alpha$ and H$\beta$ lines (hereafter, m6), and 
\item the mask used by \citet[][hereafter, VF]{2005ApJS..159..141V} 
\end{enumerate}

\noindent
 In total, our masks consist of 331 spectral intervals from six large
regions.The masks are available at Table~\ref{tab:mask} in Appendix. For each
star the best fit solution was searched for a set of six free parameters:
effective temperature \teff, surface gravity \lgg, metallicity [M/H],
microturbulence velocity \Vmic, macroturbulence velocity \Vmac, and projected
rotational velocity \vs. The final \teff, \lgg, and [M/H] values are listed in
Table~\ref{table1}. 

\subsection{Error analysis}
\label{err}

\begin{figure*}
\centering
\includegraphics[width=\textwidth]{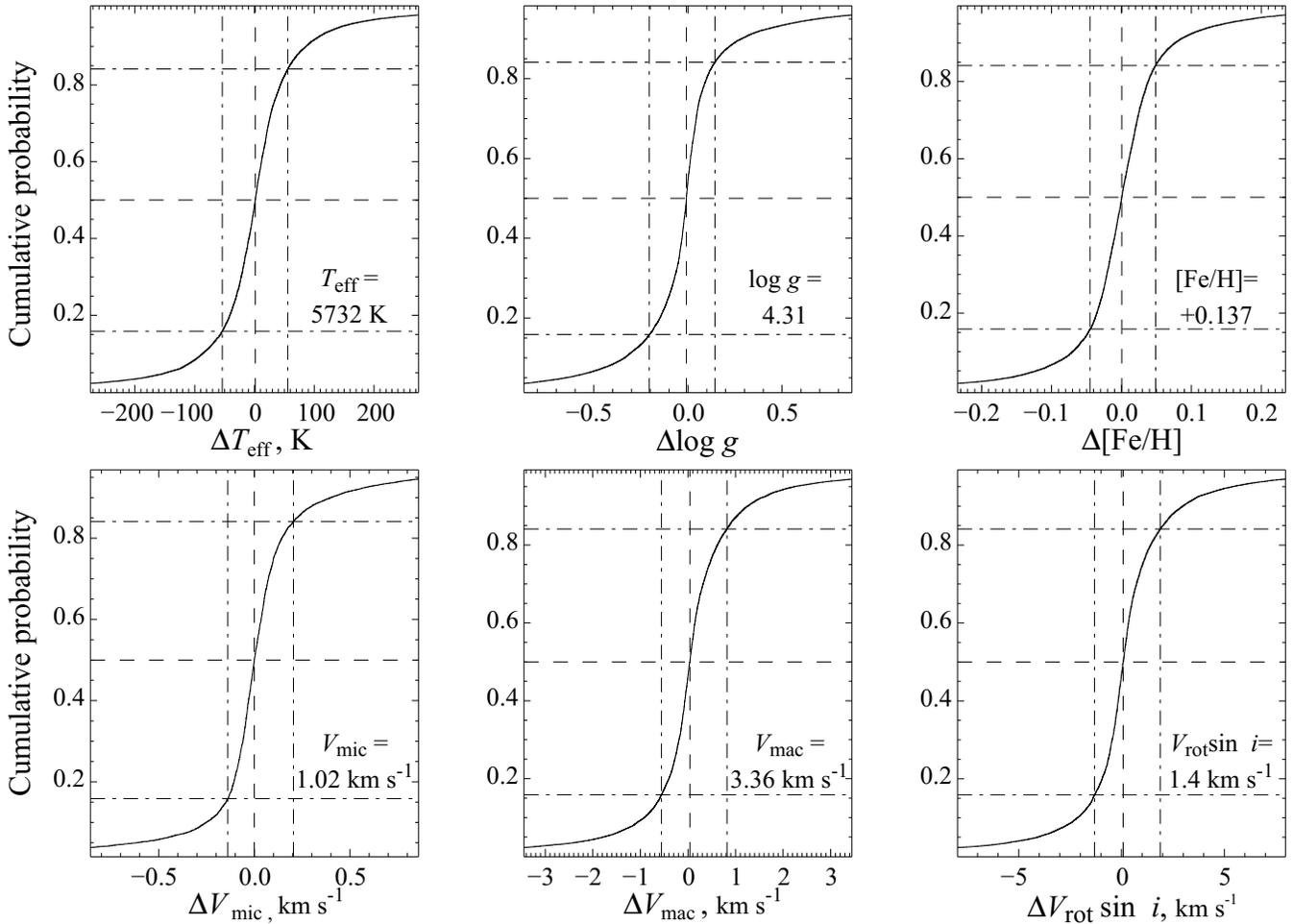}
\caption{Example of cumulative distributions obtained from SME for \teff, \lgg,
metallicity, microturbulence velocity \Vmic, macroturbulence velocity \Vmac, and
projected rotational velocity \vs\ for HD~1461 (UVES spectrum). The zero-points
of the x-axes are set at the final values obtained with SME. The horizontal dash
and dash-dotted lines show the median and the 1~$\sigma$ levels, respectively.} 
\label{cumulatives} 
\end{figure*}

\begin{figure*}
\centering
\includegraphics[width=\textwidth]{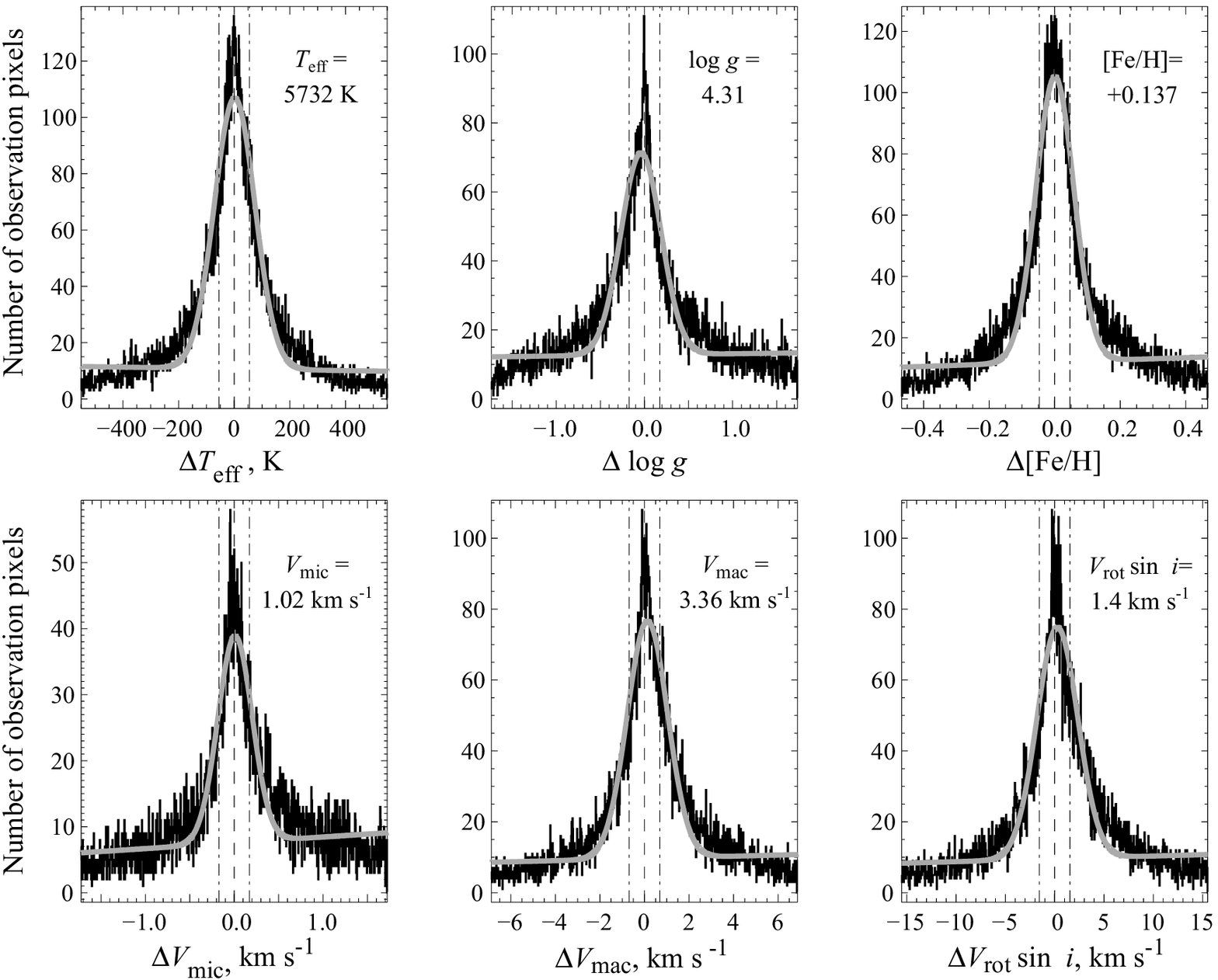}
\caption{Example of density distributions for the same parameters as in
Figure~\ref{cumulatives}. Best fit Gaussians of the central peaks are shown by
gray colour. Dashed and dash-dotted lines indicate the median and the
$\pm\sigma$ range estimated from the cumulative distributions.}
\label{histograms} 
\end{figure*}

The newly implemented method for estimating uncertainties in the free
parameters within SME is based on the fit residuals, partial derivatives, and
data uncertainties. We construct a cumulative distribution using all data pixels
for a given free parameter $p$. This distribution describes the fraction of
spectral pixels that requires a change of $\delta p$ or less to achieve a
"perfect" fit. Examples of such distributions for different parameters in
HD~1461 (UVES spectrum) are given in Figure~\ref{cumulatives}. The position of
the data pixels on the $x$-axis is estimated from the residuals and the partial
derivative of the synthetic spectrum over the particular parameter. Probability
density distributions for such uncertainty estimate have very wide wings due to
pixels insensitive to the selected parameter or pixels impossible to fit due to
erroneous observations or/and incorrect atomic and molecular data, but they do
have a clear central peak. Examples of the central part of the density
distributions for the same parameters are given in Figure~\ref{histograms}. To
guide the reader's eye we also added the best fit Gaussians and marked the
median and the $\pm\sigma$ range estimated from the cumulative distributions.

The fact that the central part is not too far from a Gaussian confirms that we
can still apply normal distribution standard deviation concept for assessing
parameter uncertainties for a bulk of observations; this is anyway more easily
done with a cumulative distribution because it does not require an a priori
knowledge of the bin size and of the clipping range for the central part. The
final parameter uncertainty includes both systematic uncertainty (model
limitations) and the observational one. While our method still ignores the
cross-talk between different parameters it shows massive improvement over the
uncertainties based on the covariance matrix (also evaluated by SME). The
uncertainties derived from the diagonal of the covariance matrix are effectively
a projection of the observational errors onto the model parameter space using
partial derivatives averaged over all data pixels. For high S/N observations the
contribution of the observational errors is negligible in comparison to model
limitations, and the resulting uncertainty estimates are unrealistically small. 
The mathematical background of the new approach can be found in the upcoming
paper by \citet{SME2015ApJ}.

Note that the estimated median values in Fig.~\ref{cumulatives} are close to,
but not exactly matching, the SME result. The reason is that the error estimate
procedure presented here is not taking into account the degeneracy/correlation
between parameters. The final estimated errors are listed in Table~\ref{table1}.
 The starting points used for the SME's optimisation algorithm were taken
close to the previously published parameters. 
We check the robustness of the convergence algorithm implemented in SME by
performing 1000 runs for the ESPaDONs spectrum of HD~69830
(mask4) with initial guesses for \teff, \lgg, and [M/H] equally spaced across
the volume: [4972:5872; 4.015:4.915; -0.4945:0.4055]. The solutions were found
to converge within 3~K in \teff, 0.005~dex in \lgg, and 0.002~dex in [M/H].

The influence of the continuum rectification on the final atmospheric
parameters was checked in the following way. For three stars of different
effective temperature and gravity, HD~9826, HD~69830, and HD~23249, we run SME
with the m4 and m6 masks using observations with the continuum rectification
independently made by three of us, i.e. T.~Ryabchikova, Yu.~Pakhomov, and
L.~Fossati; we did not find any systematic bias in our parameter
determinations caused by the different continuum placements. The differences in
the derived gravity and metallicity did not exceed 0.03~dex and 0.01~dex,
respectively. For the effective temperature, a maximum difference of 20~K was
obtained for HD~69830, but this value is smaller than the uncertainty.

\begin{figure}
\centering
\includegraphics[width=\columnwidth]{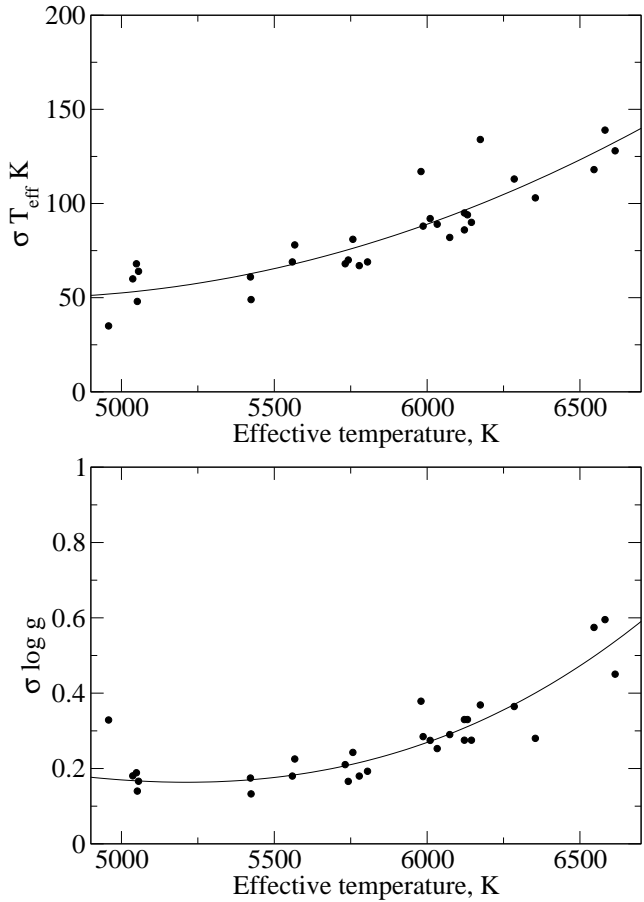}
\caption{Uncertainty in the effective temperature (top) and surface gravity
(bottom) as a function of the effective temperature. The solid line shows the
quadratic polynomial fit to the data.} 
\label{temp-errors} 
\end{figure}

A close look at the results presented in Table~\ref{table1} shows that the
spectral resolution of the observations does not significantly affect the
derived parameters: the overall difference in the effective temperature, surface
gravity, and metallicity does not exceed the error bars.
Figure~\ref{temp-errors} shows that the uncertainties in temperature and gravity
increase with increasing effective temperature. In order to consistently compare
the uncertainties for each star, all error bars have been interpolated in order
to provide the uncertainties obtained from a spectrum with S/N\,=\,200.
Figure~\ref{temp-errors} shows that a relative uncertainty in the effective
temperature changes from 1~\%\ to 2~\%\ over the 5000--6700~K temperature
range. 

\begin{figure}
\centering
\includegraphics[width=\columnwidth]{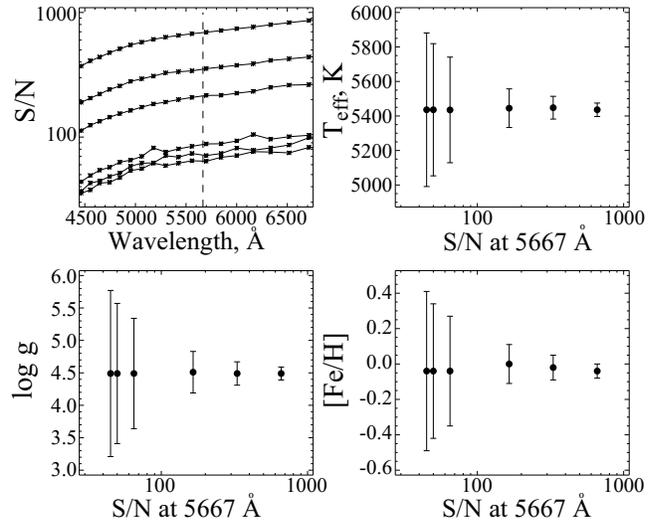}
\caption{Dependence of the inferred atmospheric parameters (\teff, \lgg, and
[M/H]) and their uncertainties on the S/N of the observed spectra. The top-left
panel gives the S/N in logarithmic scale as a function of wavelength.} 
\label{noise} 
\end{figure}

We further checked a dependence of the derived parameters and their
uncertainties on the spectrum quality (i.e., S/N). We took the ESPaDONs spectrum
of HD~69830 and introduced an additional white noise component in order to
decrease the S/N. In the SME analysis we used the common initial parameters for
each spectrum. The results are shown in Fig.~\ref{noise}. The \teff\ values
derived from different spectra agree within $\pm$20~K, while the \lgg\ values
range between 4.38 and 4.53, not centering at \lgg\ = 4.49 that was derived from
the spectrum with the highest S/N. A similar behaviour was also found for the
metallicity. For the effective temperature, surface gravity, and metallicity SME
finds solutions in a range much smaller than the typical error estimates. Our
conclusion is therefore that the S/N should not have a great influence on the
derived parameters even though the error bars increase dramatically with
decreasing S/N.

 For high S/N spectra the derived uncertainty is driven by model limitations
(systematic error). In our case these systematic errors are $\pm$20-30~K in
\teff, $\pm$0.05-0.08~dex in \lgg, and $\pm$0.02~dex in [M/H], if we consider
spectra with S/N$\geq$1000. We also checked the possible effect of the model
grids used in SME on the derived parameters. For a few, including the Sun, we
run SME with ATLAS9 and LLmodels grids of stellar atmospheres. LLmodels give
effective temperatures of 10-15~K higher than MARCS models for stars cooler than
the Sun and of 10-20~K lower for hotter stars. Surface gravity changes by less
than +0.02~dex for stars with \teff$>$5700~K and by +0.03--+0.05~dex for cooler
stars. Metallicity is practically unaffected. With ATLAS9 models the
corresponding changes are $-30$~K to +50~K in \teff\, and $-0.03$~dex or less in
surface gravity.  All these changes lie within the error limits obtained with
MARCS models. The comparison for the solar spectrum allows us to evaluate
possible offsets, because the Sun is then only star with a direct, model
independent determination of the effective temperature and surface gravity. We
obtain possible offsets of $-20$~K, $-0.03$~dex, $-0.03$~dex for effective
temperature, surface gravity and metallicity. All these values are compatible
with the corresponding errors given by spectrum fitting.
 
\section{Analysis of the derived atmospheric parameters}
\label{param}
We compare here the obtained atmospheric parameters with those present in the
literature and based on interferometric and spectroscopic methods.   

\subsection{Comparison with the interferometric parameters}
\label{interf}

\begin{figure*}
\includegraphics[width=\textwidth,clip]{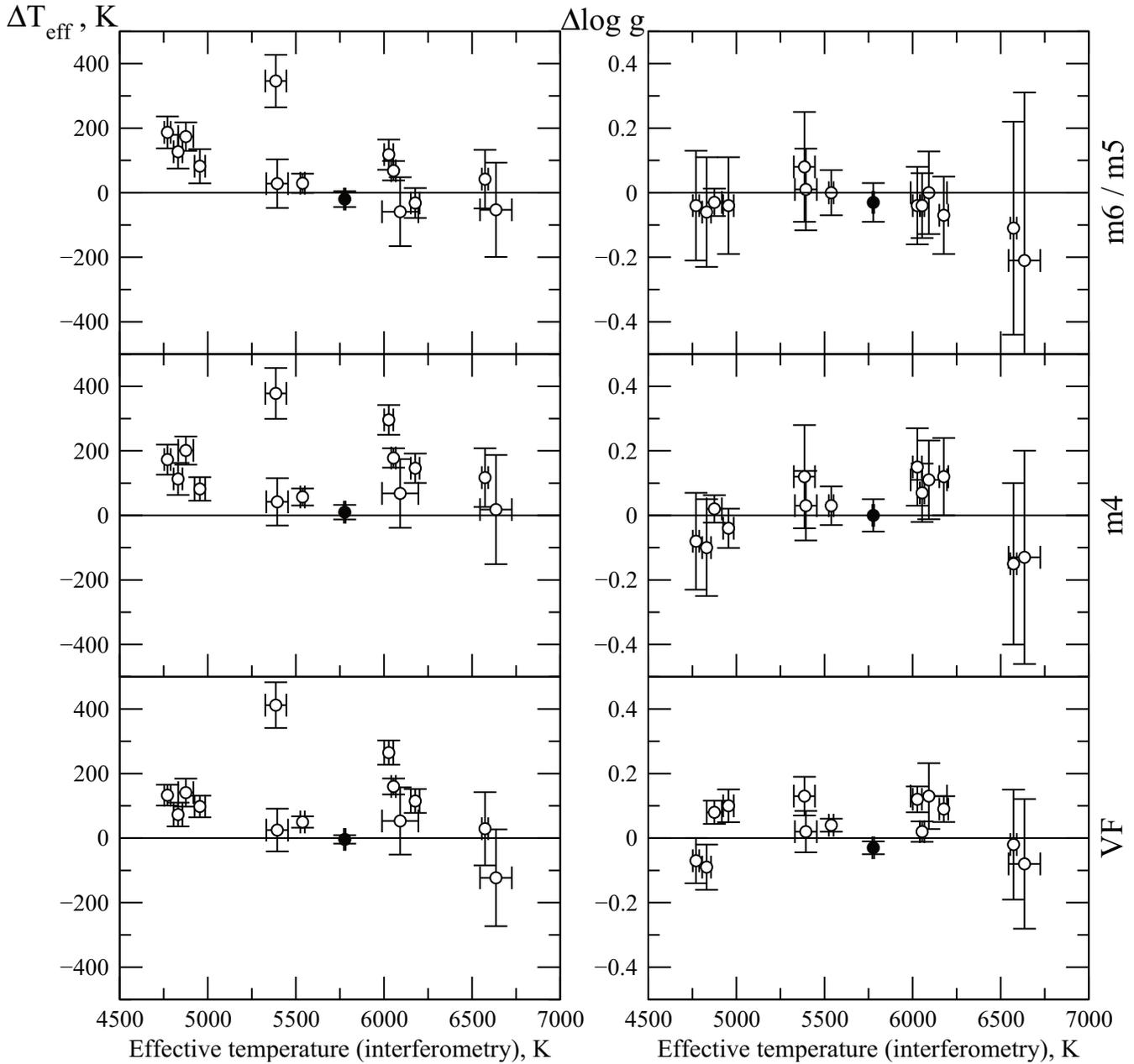}
\caption{Differences between the SME and interferometric \teff\ (left) and \lgg\
(right) values as a function of the interferometric effective temperature. The
differences are shown on the basis of the adopted masks for the SME analysis:
m6/m5 (top), m4 (middle), and VF (bottom). The Sun is shown by a filled
circle.} 
\label{diff-interf} 
\end{figure*}

From interferometric data, the effective temperature is derived from the
measured angular diameter and bolometric flux. The  interferometric \teff\ for
Procyon, HD~49933, $\upsilon$~And, $\beta$~Vir, $\delta$~Eri, and HD~103095 were
extracted from \citet[][and references therein]{2013ApJ...771...40B}. For other
stars the corresponding data were taken from \citet[][HD~209458 and
HD~189733]{2015MNRAS.447..846B}, \citet[][HD~1461 and
61~Vir]{2014MNRAS.438.2413V}, and \citet[][HD~69830]{2015ApJ...800..115T}. These
values are listed in Table~\ref{table1}. Bolometric fluxes for all stars were
calculated using the spectral energy distribution (SED) fitting code presented
by \citet{2008ApJS..176..276V}. For the solar atmosphere we used the canonical
parameters \citep{1996Sci...272.1286C}. Three stars have two or more independent
measurements of the angular diameter \citep[see][and references
therein]{2013ApJ...771...40B}.

For Procyon the angular diameter measurements lead to effective temperatures
that agree within the quoted uncertainties. For the other two stars, with quoted
uncertainties of $\sim$25~K or less, the interferometric temperatures differ by
60~K to 150~K. While the effective temperature is a measured parameter, though
indirect, the surface gravity is only inferred from the stellar radius measured
with interferometry and the star's mass, derived on the basis of evolutionary
tracks. There are two stars, HD~209458 and HD~189733, for which the surface
gravities have been derived from the binary solution obtained from the analysis
of the transit and radial velocity curves of their planets
\citep{2015MNRAS.447..846B}. We estimated the uncertainty in \lgg\ assuming a
5~\% uncertainty in the mass given by \citet{2013ApJ...771...40B} and the
quoted uncertainty in the measured radius.

Figure~\ref{diff-interf} shows the differences between the SME (using different
masks) and interferometric \teff\ and \lgg\ values as a function of the
interferometric effective temperature. For stars hotter than the Sun the closest
agreement with the interferometric parameters is achieved, when using the m6 or
m5 masks within SME, while the other two masks tend to overestimate the
effective temperature. Below 5000~K we obtain systematically higher effective
temperatures compared to those derived from interferometry, however, our results
agree well with those obtained with other spectroscopic methods (see
Sect.~\ref{classical}). The only outlier is HD~1461: we believe that its
measured angular diameter is too high resulting in lower effective temperature.
Plenty of different modern methods of parameter determination provide results
for HD~1461 which agree with each other, but not with the interferometric ones
\citep[see the PASTEL catalogue of stellar parameters;][]{2010AA...515A.111S}.
When excluding this particular star, the
average temperature differences between the SME and interferometric values are
50~K, 80~K, and 120~K for m6, VF, and m4 masks, respectively. When considering
instead only stars hotter than 5100~K, the corresponding average differences are
18~K, 60~K, and 110~K. In all comparisons the standard deviation in \teff\
varies from 60~K (m6) to 90~K (m4, VF). The difference in \lgg\ varies from
-0.04~dex (m6) to +0.04~dex (m4, VF) with a standard deviation of 0.08~dex.   

\begin{figure*}
\includegraphics[width=\textwidth]{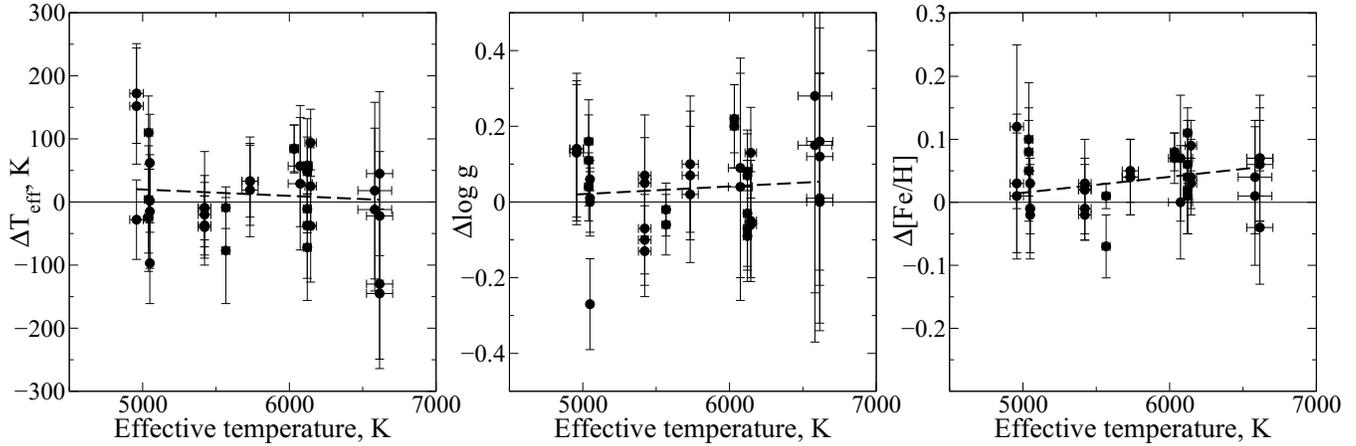}
\caption{Differences in temperature (left), gravity (center), and metallicity
(right) between the SME analysis with the m6/5 masks and the corresponding
literature spectroscopic values as a function of the SME effective temperature.
Linear regressions weighted by the uncertainties are indicated by dashed
lines.} 
\label{m6-spectr}
\end{figure*}

\begin{figure*}
\includegraphics[width=\textwidth]{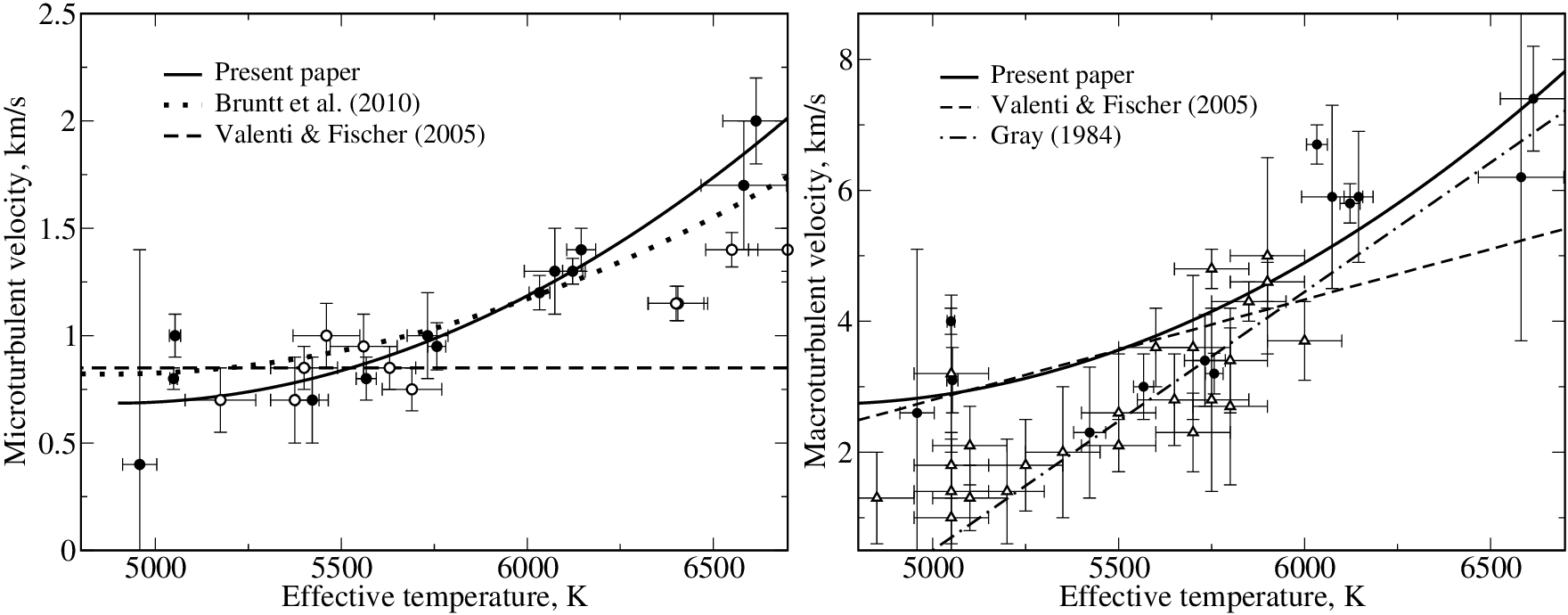}
\caption{Dependence of \Vmic\ (left panel) and \Vmac\ (right panel) on the
effective temperature. The results of the present paper are shown by filled
circles. The quadratic polynomial fits of our results are shown by solid lines
in both panels. Open circles show the \Vmic\ values from
\citet{2013MNRAS.428.3164D}, while open triangles show the \Vmac\ data from
\citet{1997MNRAS.284..803S}. The sources for analytical approximations are
quoted in each panel}. 
\label{V-turb}
\end{figure*}

\subsection{Comparison with other spectroscopic studies}
\label{classical}

Table~\ref{table1} lists the atmospheric parameters derived in the literature
using various spectroscopic methods, together with their sources, and
Fig.~\ref{m6-spectr} displays the differences in temperature, gravity, and
metallicity between the SME analysis with the m6/m5 masks and the corresponding
literature data. No significant trends and offsets are found. Using the VF
and m4 masks leads to an overestimation of all three parameters for the hottest
stars. Instead, using the mask including the hydrogen lines (m6/m5)
removes this offset. 
 
The average difference between the stellar parameters derived with SME (m6/m5
masks) and other spectroscopic methods was found to be 12$\pm$66~K for the
effective temperature, 0.04$\pm$0.12~dex for the surface gravity, and
0.04$\pm$0.04~dex for the metallicity. It may be considered as
method-to-method dispersion. The dispersion values corresponds to the error
estimates derived by SME for S/N=200.

\subsection{Micro- and macroturbulent velocities}

When deriving atmospheric parameters for a large number of stars by means of a
spectrum fitting technique, both micro and macroturbulent velocities (\Vmic\ and
\Vmac) are usually not directly estimated, but are either fixed \citep[e.g.,
\Vmic\ in][]{2005ApJS..159..141V} or estimated on the basis of analytical
approximations \citep[e.g., \Vmac\ in][]{2012ApJ...757..161T}. Our SME analysis
allowed us to derive both these parameters, with the corresponding error bars,
for each star. Figure~\ref{V-turb} shows stellar \Vmic\ and \Vmac\ parameters as
a function of the effective temperature.

In general, our measurements agree with those given by previous studies. The
analytical expression by \citet{2010MNRAS.405.1907B} for \Vmic\ seems to fit our
data in a reasonable way, while using a fixed value of \Vmic\ would be
inappropriate for stars hotter than 6000~K. The analytical formula proposed by
\citet{2005ApJS..159..141V} underestimates \Vmac\ for the hottest stars, while
it reasonably represents our measurements for stars cooler than 5500~K.
The \citet{1984ApJ...281..719G} formula seems to be a good approximation for
\Vmac\ in stars hotter than the Sun.   

\section{Abundances}
\label{spectr}

We test the atmospheric parameters obtained with the m6 mask by inspecting
abundances from two ionisation stages for Ti and Fe and from the atomic
\ion{C}{i} and molecular CH species. The elements Ti and Fe were chosen because
of the large number of lines with precise laboratory atomic data available at
optical wavelengths. For \ion{Ti}{i} and \ion{Ti}{ii} we used the homogeneous
set of laboratory transition probabilities by \citet{LGWSC} and \citet{WLSC},
while for \ion{Fe}{i} and \ion{Fe}{ii} the line parameters were extracted from
the third version of the VALD database \citep{2015PhyS...90e4005R}. We also
considered the C abundance derived from atomic \ion{C}{i} and molecular CH lines
that are known to be sensitive to \teff\ variations. Line parameters were also
extracted from VALD. While the C$_2$ Swan system lines were included in fitting
procedure, for abundances we employed the CH lines in 4200-4400~\AA\ region
because these lines are strong and are easily measured in spectra of all stars
of our program. Unblended C$_2$ lines suitable for abundance determinations
practically disappeared in spectra of stars hotter than 6000~K. A list of the
lines used for the abundance determination, together with the adopted line
parameters, is given in Table~\ref{linelist} (online material). In total we used
8 lines of \ion{C}{i} in the visible and near IR regions, 8 bands of CH, 21
lines of \ion{Ti}{i}, 10 lines of \ion{Ti}{ii}, 73 lines of \ion{Fe}{i}, and 28
lines of \ion{Fe}{ii}. 

Element abundances from the atomic lines were calculated based on the non-local
thermodynamic equilibrium (NLTE) line formation. To solve the coupled radiative
transfer and statistical equilibrium (SE) equations, we used a revised version
of the {\sc detail} code \citep{detail}. The update was described by
\citet{mash_fe}. In the atmospheric parameter range covered by the analysed
stars, departures from LTE are negligible for \ion{Ti}{ii} and \ion{Fe}{ii}
lines and small for lines of \ion{Ti}{i} and \ion{Fe}{i} and lines of \ion{C}{i}
in the visible spectral range. The molecular CH lines are considered to be free
of NLTE effects. For \ion{Ti}{i} and \ion{Ti}{ii} the NLTE calculations were
performed using a comprehensive model atom constructed by \citet{ti_atom}. For
cool stars the main source of uncertainty in NLTE calculations for \ion{Ti}{i}
(as well as for \ion{Fe}{i}) is poorly known inelastic collision with hydrogen
atoms. In this study they are treated employing the Drawinian
\citep{1968ZPhy..211..404D,1969ZPhy..225..483D} rates scaled by a factor of
\kH\,=\,0.5. It is worth noting that for HD~49933 (\teff\,=\,6580~K; \lgg\,=\,4.
0; [M/H]\,=\,$-$0.48~dex) using \kH\,=\,0.5 leads to 0.019~dex higher average
abundance from the \ion{Ti}{i} lines compared to that for \kH\,=\,1.0. Effect of
varying \kH\ is even smaller for the Sun, with a difference of 0.008~dex between
applying \kH\,=\,0.5 and 1.0.  For iron, we used a comprehensive
\ion{Fe}{i}-\ion{Fe}{ii} model atom treated by \citet{mash_fe} and employed
\kH\,=\,0.5, as deduced by \citep{2015ApJ...808..148S} from analysis of the
nearby dwarf stellar sample. The NLTE calculations for the \ion{C}{i} lines were
performed with the method from \citet{Alexeeva_carbon}.

\begin{table}
\caption{Atomic and molecular line parameters} 
\label{linelist}
\begin{tabular}{|c|c|c|c|l|}
\hline
Wavelength,\AA & Ion & $E_{i}$, eV & log($gf$) & Reference \\ 
\hline
\hline
  4218.7130 &   CH    &   0.411 &  -1.339 &   1                  \\
  4218.7340 &   CH    &   0.411 &  -1.361 &   1                  \\
   ...      &   ...   &   ...   &   ...   &   ...  \\  
  4489.7391 &   \ion{Fe}{i}  &   0.121 &  -3.966 &   2                  \\
  4491.3971 &   \ion{Fe}{ii} &   2.856 &  -2.700 &   3                   \\
  4493.5220 &   \ion{Ti}{ii} &   1.080 &  -2.780 &   4                   \\
    ...     &   ...   &  ...    &  ...    &  ...                  \\
\hline
\end{tabular}
\\
\begin{flushleft}
References: (1) \citet{JLIY}; (2) \citet{FMW}; (3) \citet{KK}; (4) \citet{WLSC};
(5) \citet{RU}; (6) \citet{LGWSC}; (7) \citet{T83av};, (8) \citet{BGHR};
(9) \citet{Kurucz}; (10) \citet{BWL}; (11) \citet{NIST10}; (12) \citet{BSScor}
(corrected); (13) \citet{BKK}; (14)\citet{HLGN}; (15) \citet{BK}; (16)
\citet{MRW};
(17) \citet{2002ApJ...573L.137A}.\\
(This table is available in its entirety in a machine-readable form in the
online
journal. A portion is shown here for guidance regarding its form and content.)
\end{flushleft}
\end{table}

The departure coefficients (the ratios of NLTE to LTE level populations)
calculated for each atmospheric model with the \textsc{detail} code were then
implemented in the \textsc{synthV\_NLTE} code. This software, presented in
\citet{synthV}, calculates the spectrum emerging from the static, 1-D model
atmosphere, and it was tuned for the modelling of early B- to late M-type stars.
The code was originally employing the LTE approximation, and in this study it
was modified to take into account the pre-computed departure coefficients for
various chemical species. We further integrated it within the IDL
\textsc{binmag3} code\footnote{http://www.astro.uu.se/$\sim$oleg/download.html},
written by O. Kochukhov, finally allowing us to determine the best fit to the
observed line profiles with the inclusion of the NLTE effects. The spectrum
synthesis code adopted by SME and the \textsc{synthV\_NLTE} code use different
algorithms for the radiative transfer solution, which may result in systematic
differences in the derived abundances. We compared the two codes as follows. We
used the synthetic spectrum computed by SME for HD~69830 (ESPADONS spectrum) in
the 5640-50~\AA\ wavelength region as the ``observed spectrum`` for
(\textsc{synthV\_NLTE} + \textsc{binmag3}). In the latter code, abundances of
eight elements listed in Table~\ref{tab:cmp_abund} were allowed to vary, when
fitting the SME 'observed spectrum'. Table~\ref{tab:cmp_abund} shows the LTE
abundances from calculations with SME and \textsc{synthV\_NLTE} +
\textsc{binmag3}. The SME code returns an overall element abundance based on
all the lines of each given element in the SME spectral windows, while
\textsc{synthV\_NLTE} + \textsc{binmag3} provides abundances from individual
lines, which were averaged for each given element. It can be seen that the
two different spectrum synthesis codes used in our study provide consistent
abundances within 0.02~dex.

\begin{table}
\caption{LTE abundances derived by SME for HD~69830 (second column) and by {\sc
synthV\_NLTE} + {\sc binmag3} for the SME 'observed spectrum' (third column).
See text for more details.} 
\label{tab:cmp_abund}
\centering
\begin{tabular}{|l|c|c|}
\hline
\header{Element} & \multicolumn{2}{c}{Abundance $\log(N_{el}/N_{tot}$)}      \\
\cline{2-3}
   &   \header{SME}     & \header{synthV\_NLTE+} \\ 
   &                    & \header{binmag} (LTE) \\ 
\hline
Si & -4.57(0.17) &  -4.56 \\
Sc & -8.87(0.04) &  -8.87 \\
Ti & -7.11(0.07) &  -7.11 \\
V  & -8.11(0.06) &  -8.10 \\
Cr & -6.44(0.07) &  -6.42 \\
Fe & -4.60(0.06) &  -4.59 \\
Co & -7.23(0.10) &  -7.22 \\          
Ni & -5.87(0.05) &  -5.86 \\
\hline
\end{tabular}
\end{table}

\begin{figure}
\includegraphics[width=\columnwidth,clip]{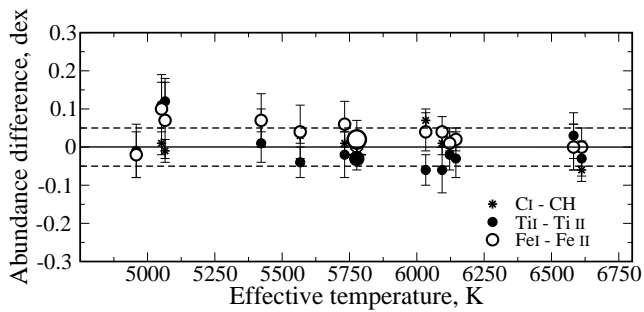}
\caption{Abundance differences between the atomic \ion{C}{i} and molecular CH
lines (stars), between lines of \ion{Ti}{i} and \ion{Ti}{ii} (filled circles),
and between lines of \ion{Fe}{i} and  \ion{Fe}{ii} (open circles). The solar
values are shown by larger size symbols. The solid line shows the perfect
ionisation balance, while the dashed lines indicate the $\sigma\,=\,\pm$0.05~dex
level.} 
\label{equilibr} 
\end{figure}

Stellar abundances of C, Ti, and Fe calculated with the adopted atmospheric
models using {\sc synthV\_NLTE} + {\sc binmag3} are presented in
Table~\ref{abund}. The abundance differences between the two ionisation stages
for Ti and Fe and between the atomic \ion{C}{i} and molecular CH species are
shown in Fig.~\ref{equilibr}. We consider $\pm$0.05~dex to be the typical error
bars for our abundance determination.  Such an uncertainty corresponds to that
obtained by a variation of $\pm$0.1~dex in the surface gravity and $\pm$50--70~K
in the effective temperature for solar-type stars. This agrees with the results
presented in Sect.~\ref{interf} and \ref{classical}. For all the stars, except
HD~189733 and HD~23249, the ionisation equilibria are achieved within 0.05~dex.

\begin{table*}
\caption{NLTE abundances of C, Ti, and Fe derived from the atomic lines and the
LTE abundances from the molecular CH lines together with their standard
deviations given in parentheses.}
\renewcommand\arraystretch{1.3}
\vspace{2mm}
\begin{large}
\centering
\vspace{5mm}
\label{abund}
\setlength{\tabcolsep}{1mm}

\begin{tabular}{|l|c|c|c|c|c|c|c|c|}
\hline
\headerb{Name/HD} &  \header{\teff, K} & \header{lg~$g$} &
\multicolumn{6}{|c|}{Abundance $\log(N_{el}/N_{tot}$)} \\
\cline{4-9}
       &      &      &  \header{CH} & \header{\ion{C}{i}} & \header{\ion{Ti}{i}}
& \header{\ion{Ti}{ii}} & \header{\ion{Fe}{i}} & \header{\ion{Fe}{ii}} \\
\hline
Sun       & 5777      &	4.44     &  $-$3.60(02)  & $-$3.62(01) & $-$7.09(03) &
$-$7.06(04) & $-$4.55(05) &  $-$4.57(06) \\
Procyon   & 6615      & 3.89     &  $-$3.59(02)  & $-$3.66(01) & $-$7.12(06) &
$-$7.09(08) & $-$4.57(05) &  $-$4.57(08) \\
HD 49933  & 6582      & 4.00     &  $-$4.09(05)  & $-$4.09(02) & $-$7.52(06) &
$-$7.55(04) & $-$5.08(06) &  $-$5.08(06) \\
HD 9826   & 6145      & 4.06     &  $-$3.62(01)  & $-$3.59(02) & $-$7.04(05) &
$-$7.01(06) & $-$4.48(07) &  $-$4.50(06) \\
HD 102870 & 6122      & 4.07     &  $-$3.53(03)  & $-$3.51(03) & $-$6.96(04) &
$-$6.94(05) & $-$4.40(06) &  $-$4.41(06) \\
HD 149026 & 6074      & 4.18     &  $-$3.42(03)  & $-$3.41(02) & $-$6.87(06) &
$-$6.81(06) & $-$4.27(06) &  $-$4.31(06) \\
HD 209458 & 6033      & 4.28     &  $-$3.78(02)  & $-$3.71(02) & $-$7.14(04) &
$-$7.08(05) & $-$4.58(06) &  $-$4.62(05) \\
HD 1461   & 5732      & 4.31     &  $-$3.50(02)  & $-$3.49(02) & $-$6.95(06) &
$-$6.94(06) & $-$4.38(06) &  $-$4.43(05) \\
HD 115617 & 5567      & 4.42     &  $-$3.72(02)  & $-$3.71(03) & $-$7.09(04) &
$-$7.05(06) & $-$4.58(07) &  $-$4.62(06) \\
HD 69830  & 5422      & 4.47     &  $-$3.72(02)  & $-$3.66(05) & $-$7.07(05) &
$-$7.08(06) & $-$4.57(07) &  $-$4.63(06) \\
HD 189733 & 5065      & 4.55     &  $-$3.74(03)  & $-$3.75(04) & $-$6.98(06) &
$-$7.10(06) & $-$4.53(10) &  $-$4.60(09) \\
HD 23249  & 5052      & 3.76     &  $-$3.61(03)  & $-$3.60(03) & $-$6.89(07) &
$-$7.00(06) & $-$4.45(09) &  $-$4.55(07) \\
HD 103095 & 4958      & 4.52     &  $-$5.17(05)  &             & $-$8.23(08) &
$-$8.22(06) & $-$5.94(06) &  $-$5.92(05) \\
\hline
\end{tabular}
\end{large}
\end{table*}

For HD~189733 and HD~23249 their carbon abundances derived from the atomic and
molecular lines are consistent, while the \ion{Ti}{i}--\ion{Ti}{ii} and
\ion{Fe}{i}--\ion{Fe}{ii} abundance differences reach +0.10~dex. These positive
differences may indicate an overestimation of the effective temperature as
derived by SME and support the lower effective temperature from interferometry.
We inspected abundances of C, Ti, and Fe calculated with the interferometric
atmospheric parameters, and we obtained an abundance difference of $\sim
-0.10$~dex for \ion{Ti}{i}--\ion{Ti}{ii} and \ion{Fe}{i}--\ion{Fe}{ii} and a
very large difference of 0.4~dex between \ion{C}{i} and CH. Most likely, for
both stars their effective temperatures lie in between the SME and the
interferometric values.    

\section{Conclusions}
\label{concl}

We used a sample of well studied F-, G-, and K-type stars to perform an
extensive test of the accuracy of atmospheric parameters derived with the
fitting procedure SME, widely applied in the literature for late-type stars. In
our analysis we used high-quality and high-resolution spectra obtained with a
wide range of different instruments and telescopes. We adopted three different
masks for the spectral analysis and implemented a new, more accurate scheme for
the determination of the uncertainties in the atmospheric parameters. We would
like to emphasise that our approach provides self-consistent spectroscopic
results without any external constraints on any of five atmospheric parameters:
\teff, \lgg, [M/H], \Vmic, and \Vmac.

 We compared our results with those obtained with interferometry arriving at the
following conclusions.

\begin{itemize}     
\item Atmospheric parameters derived with SME on the basis of different masks
agree within the uncertainties for stars cooler than 5500~K, while for hotter
stars we recommend to include the \hb\ and/or \ha\ lines to the spectrum fitting
procedure.
\item The uncertainty in the effective temperature is 50--70~K for the
S/N\,=\,200 spectra of the main-sequence F-, G-, K-type stars.
\item Given the spectral ranges, spectral resolution, and S/N values explored
here, we were unable to measure surface gravity with an accuracy of \lgg\ better
than 0.1~dex using SME.
\item The typical uncertainty in the metallicity derived with SME is
0.05-0.06~dex. 
\end{itemize}

\section*{Acknowledgements}
This work was supported by the Federal Agency for Science and Innovations (grant
No.~8529), the Russian Foundation for Basic Research (grant 15-02-06046). T.S.
and S.A. thank the RFBR grant 14-02-31780. L.M. and T.S. are thankful to the
Swiss National Science Foundation (SCOPES project No. IZ73Z0-152485) for partial
financial support. This study is based on observations made with ESO Telescopes
at the La Silla Paranal Observatory under programme ID 085.D-0124 and
MegaPrime/MegaCam, a joint project of CFHT and CEA/IRFU, at the CFHT, which is
operated by the National Research Council (NRC) of Canada, the Institut National
des Science de l'Univers of the Centre National de la Recherche Scientifique
(CNRS) of France, and the University of Hawaii. We employed the data products
produced at Terapix available at the Canadian Astronomy Data Centre as part of
the Canada-France-Hawaii Telescope Legacy Survey, a collaborative project of NRC
and CNRS. This research has made use of the Keck Observatory Archive (KOA),
which is operated by the W. M. Keck Observatory and the NASA Exoplanet Science
Institute (NExScI), under contract with the National Aeronautics and Space
Administration. The Observatory was made possible by the generous financial
support of the W.M. Keck Foundation. We thank anonymous referee for the
insightful remarks.

\bibliographystyle{mn2e}
\bibliography{ref}

\begin{thebibliography}{}

\bibitem[\protect\citeauthoryear{{Alexeeva} \& {Mashonkina}}{{Alexeeva} \&
  {Mashonkina}}{2015}]{Alexeeva_carbon}
{Alexeeva} S.~A.,  {Mashonkina} L.~I.,  2015, \mnras, 453, 1619

\bibitem[\protect\citeauthoryear{{Allende Prieto}, {Lambert} \&
  {Asplund}}{{Allende Prieto} et~al.}{2002}]{2002ApJ...573L.137A}
{Allende Prieto} C.,  {Lambert} D.~L.,    {Asplund} M.,  2002, \apjl, 573, L137

\bibitem[\protect\citeauthoryear{{Bagnulo}, {Jehin}, {Ledoux}, {Cabanac},
  {Melo}, {Gilmozzi} \& {ESO Paranal Science Operations Team}}{{Bagnulo}
  et~al.}{2003}]{2003Msngr.114...10B}
{Bagnulo} S.,  {Jehin} E.,  {Ledoux} C.,  {Cabanac} R.,  {Melo} C.,  {Gilmozzi}
  R.,    {ESO Paranal Science Operations Team} 2003, The Messenger, 114, 10

\bibitem[\protect\citeauthoryear{{Bard}, {Kock} \& {Kock}}{{Bard}
  et~al.}{1991}]{BKK}
{Bard} A.,  {Kock} A.,    {Kock} M.,  1991, \aap, 248, 315

\bibitem[\protect\citeauthoryear{{Bard} \& {Kock}}{{Bard} \& {Kock}}{1994}]{BK}
{Bard} A.,  {Kock} M.,  1994, \aap, 282, 1014

\bibitem[\protect\citeauthoryear{{Barklem}, {Piskunov} \& {O'Mara}}{{Barklem}
  et~al.}{2000}]{2000A&A...363.1091B}
{Barklem} P.~S.,  {Piskunov} N.,    {O'Mara} B.~J.,  2000, \aap, 363, 1091

\bibitem[\protect\citeauthoryear{{Baschek}, {Garz}, {Holweger} \&
  {Richter}}{{Baschek} et~al.}{1970}]{BGHR}
{Baschek} B.,  {Garz} T.,  {Holweger} H.,    {Richter} J.,  1970, \aap, 4, 229

\bibitem[\protect\citeauthoryear{{Blackwell}, {Shallis} \&
  {Simmons}}{{Blackwell} et~al.}{1980}]{BSScor}
{Blackwell} D.~E.,  {Shallis} M.~J.,    {Simmons} G.~J.,  1980, Astron. and
  Astrophys., 81, 340

\bibitem[\protect\citeauthoryear{{Blanco-Cuaresma}, {Soubiran}, {Jofr{\'e}} \&
  {Heiter}}{{Blanco-Cuaresma} et~al.}{2014}]{2014A&A...566A..98B}
{Blanco-Cuaresma} S.,  {Soubiran} C.,  {Jofr{\'e}} P.,    {Heiter} U.,  2014,
  \aap, 566, A98

\bibitem[\protect\citeauthoryear{{Boyajian}, {von Braun}, {Feiden}, {Huber},
  {Basu} \& {et al.,}}{{Boyajian} et~al.}{2015}]{2015MNRAS.447..846B}
{Boyajian} T.,  {von Braun} K.,  {Feiden} G.~A.,  {Huber} D.,  {Basu} S.,
  {et al.,} 2015, \mnras, 447, 846

\bibitem[\protect\citeauthoryear{{Boyajian}, {von Braun}, {van Belle},
  {Farrington}, {Schaefer} \& {et al.,}}{{Boyajian}
  et~al.}{2013}]{2013ApJ...771...40B}
{Boyajian} T.~S.,  {von Braun} K.,  {van Belle} G.,  {Farrington} C.,
  {Schaefer} G.,    {et al.,} 2013, \apj, 771, 40

\bibitem[\protect\citeauthoryear{{Brooke}, {Bernath}, {Schmidt} \&
  {Bacskay}}{{Brooke} et~al.}{2013}]{BBSB}
{Brooke} J.~S.~A.,  {Bernath} P.~F.,  {Schmidt} T.~W.,    {Bacskay} G.~B.,
  2013, \jqsrt, 124, 11

\bibitem[\protect\citeauthoryear{{Bruntt}}{{Bruntt}}{2009}]{2009AA...506..235B}
{Bruntt} H.,  2009, \aap, 506, 235

\bibitem[\protect\citeauthoryear{{Bruntt}, {Bedding}, {Quirion}, {Lo Curto},
  {Carrier} \& {et al.,}}{{Bruntt} et~al.}{2010}]{2010MNRAS.405.1907B}
{Bruntt} H.,  {Bedding} T.~R.,  {Quirion} P.-O.,  {Lo Curto} G.,  {Carrier} F.,
     {et al.,} 2010, \mnras, 405, 1907

\bibitem[\protect\citeauthoryear{{Buchhave}, {Latham}, {Johansen}, {Bizzarro},
  {Torres} \& {et al.,}}{{Buchhave} et~al.}{2012}]{2012Natur.486..375B}
{Buchhave} L.~A.,  {Latham} D.~W.,  {Johansen} A.,  {Bizzarro} M.,  {Torres}
  G.,    {et al.,} 2012, \nat, 486, 375

\bibitem[\protect\citeauthoryear{{Butler} \& {Giddings}}{{Butler} \&
  {Giddings}}{1985}]{detail}
{Butler} K.,  {Giddings} J.,  1985, Newsletter on the analysis of astronomical
  spectra, No. 9, University of London

\bibitem[\protect\citeauthoryear{{Castelli} \& {Kurucz}}{{Castelli} \&
  {Kurucz}}{2004}]{2004astro.ph..5087C}
{Castelli} F.,  {Kurucz} R.~L.,  2004, in {Piskunov} N.,  { Weiss} W.,   {Gray}
  D.,  eds, Modelling of Stellar Atmospheres Proceedings of the IAU Symp. No
  210, {New Grids of ATLAS9 Model Atmospheres}

\bibitem[\protect\citeauthoryear{{Cayrel}, {van't Veer-Menneret}, {Allard} \&
  {Stehl{\'e}}}{{Cayrel} et~al.}{2011}]{2011A&A...531A..83C}
{Cayrel} R.,  {van't Veer-Menneret} C.,  {Allard} N.~F.,    {Stehl{\'e}} C.,
  2011, \aap, 531, A83

\bibitem[\protect\citeauthoryear{{Christensen-Dalsgaard}, {Dappen}, {Ajukov},
  {Anderson}, {Antia} \& {et al.,}}{{Christensen-Dalsgaard}
  et~al.}{1996}]{1996Sci...272.1286C}
{Christensen-Dalsgaard} J.,  {Dappen} W.,  {Ajukov} S.~V.,  {Anderson} E.~R.,
  {Antia} H.~M.,    {et al.,} 1996, Science, 272, 1286

\bibitem[\protect\citeauthoryear{{de la Cruz Rodr\'{i}guez} \& {Piskunov}}{{de
  la Cruz Rodr\'{i}guez} \& {Piskunov}}{2013}]{2013ApJ...764...33D}
{de la Cruz Rodr\'{i}guez} J.,  {Piskunov} N.,  2013, \apj, 764, 33

\bibitem[\protect\citeauthoryear{{Doyle}, {Smalley}, {Maxted}, {Anderson},
  {Cameron} \& {et al.,}}{{Doyle} et~al.}{2013}]{2013MNRAS.428.3164D}
{Doyle} A.~P.,  {Smalley} B.,  {Maxted} P.~F.~L.,  {Anderson} D.~R.,  {Cameron}
  A.~C.,    {et al.,} 2013, \mnras, 428, 3164

\bibitem[\protect\citeauthoryear{{Drawin}}{{Drawin}}{1968}]{1968ZPhy..211..404%
D}
{Drawin} H.-W.,  1968, Zeitschrift fur Physik, 211, 404

\bibitem[\protect\citeauthoryear{{Drawin}}{{Drawin}}{1969}]{1969ZPhy..225..483%
D}
{Drawin} H.~W.,  1969, Zeitschrift fur Physik, 225, 483

\bibitem[\protect\citeauthoryear{{Fuhr}, {Martin} \& {Wiese}}{{Fuhr}
  et~al.}{1988}]{FMW}
{Fuhr} J.~R.,  {Martin} G.~A.,    {Wiese} W.~L.,  1988, Journal of Physical and
  Chemical Reference Data, Volume 17, Suppl.~4.~New York: American Institute of
  Physics (AIP) and American Chemical Society, 1988, 17

\bibitem[\protect\citeauthoryear{{Fuhrmann}}{{Fuhrmann}}{1998}]{1998AA...338..%
161F}
{Fuhrmann} K.,  1998, \aap, 338, 161

\bibitem[\protect\citeauthoryear{{Fuhrmann}}{{Fuhrmann}}{2008}]{2008MNRAS.384.%
.173F}
{Fuhrmann} K.,  2008, \mnras, 384, 173

\bibitem[\protect\citeauthoryear{{Fuhrmann}, {Axer} \& {Gehren}}{{Fuhrmann}
  et~al.}{1993}]{1993A&A...271..451F}
{Fuhrmann} K.,  {Axer} M.,    {Gehren} T.,  1993, \aap, 271, 451

\bibitem[\protect\citeauthoryear{{Gilmore}, {Randich}, {Asplund}, {Binney},
  {Bonifacio}, {Drew}, {Feltzing}, {Ferguson}, {Jeffries}, {Micela} \& et
  al.}{{Gilmore} et~al.}{2012}]{2012Msngr.147...25G}
{Gilmore} G.,  {Randich} S.,  {Asplund} M.,  {Binney} J.,  {Bonifacio} P.,
  {Drew} J.,  {Feltzing} S.,  {Ferguson} A.,  {Jeffries} R.,  {Micela} G.,
  et al. 2012, The Messenger, 147, 25

\bibitem[\protect\citeauthoryear{{Gonzalez}, {Carlson} \& {Tobin}}{{Gonzalez}
  et~al.}{2010}]{2010MNRAS.403.1368G}
{Gonzalez} G.,  {Carlson} M.~K.,    {Tobin} R.~W.,  2010, \mnras, 403, 1368

\bibitem[\protect\citeauthoryear{{Gray}}{{Gray}}{1984}]{1984ApJ...281..719G}
{Gray} D.~F.,  1984, \apj, 281, 719

\bibitem[\protect\citeauthoryear{{Gustafsson}, {Edvardsson}, {Eriksson},
  {J{\o}rgensen}, {Nordlund} \& {et al.,}}{{Gustafsson}
  et~al.}{2008}]{2008AA...486..951G}
{Gustafsson} B.,  {Edvardsson} B.,  {Eriksson} K.,  {J{\o}rgensen} U.~G.,
  {Nordlund} {\AA}.,    {et al.,} 2008, \aap, 486, 951

\bibitem[\protect\citeauthoryear{{Hannaford}, {Lowe}, {Grevesse} \&
  {Noels}}{{Hannaford} et~al.}{1992}]{HLGN}
{Hannaford} P.,  {Lowe} R.~M.,  {Grevesse} N.,    {Noels} A.,  1992, \aap, 259,
  301

\bibitem[\protect\citeauthoryear{{Heiter}, {Jofr{\'e}}, {Gustafsson}, {Korn},
  {Soubiran} \& {Th{\'e}venin}}{{Heiter} et~al.}{2015}]{2015arXiv150606095H}
{Heiter} U.,  {Jofr{\'e}} P.,  {Gustafsson} B.,  {Korn} A.~J.,  {Soubiran} C.,
    {Th{\'e}venin} F.,  2015, ArXiv Astrophysics e-prints, astro-ph/1506.06095

\bibitem[\protect\citeauthoryear{{Howard}, {Johnson}, {Marcy}, {Fischer},
  {Wright} \& {et al.,}}{{Howard} et~al.}{2010}]{2010ApJ...721.1467H}
{Howard} A.~W.,  {Johnson} J.~A.,  {Marcy} G.~W.,  {Fischer} D.~A.,  {Wright}
  J.~T.,    {et al.,} 2010, \apj, 721, 1467

\bibitem[\protect\citeauthoryear{{Jorgensen}, {Larsson}, {Iwamae} \&
  {Yu}}{{Jorgensen} et~al.}{1996}]{JLIY}
{Jorgensen} U.~G.,  {Larsson} M.,  {Iwamae} A.,    {Yu} B.,  1996, \aap, 315,
  204

\bibitem[\protect\citeauthoryear{{Kroll} \& {Kock}}{{Kroll} \&
  {Kock}}{1987}]{KK}
{Kroll} S.,  {Kock} M.,  1987, Astron. and Astrophys. Suppl. Ser., 67, 225

\bibitem[\protect\citeauthoryear{{Kurucz}}{{Kurucz}}{2015}]{Kurucz}
{Kurucz} R.~L., , 2015, On-line database of observed and predicted atomic
  transitions, http://kurucz.harvard.edu/atoms

\bibitem[\protect\citeauthoryear{{Kurucz}, {Furenlid}, {Brault} \&
  {Testerman}}{{Kurucz} et~al.}{1984}]{1984sfat.book.....K}
{Kurucz} R.~L.,  {Furenlid} I.,  {Brault} J.,    {Testerman} L.,  1984, {Solar
  flux atlas from 296 to 1300 nm}.
New Mexico: National Solar Observatory

\bibitem[\protect\citeauthoryear{{Lawler}, {Guzman}, {Wood}, {Sneden} \&
  {Cowan}}{{Lawler} et~al.}{2013}]{LGWSC}
{Lawler} J.~E.,  {Guzman} A.,  {Wood} M.~P.,  {Sneden} C.,    {Cowan} J.~J.,
  2013, \apjs, 205, 11

\bibitem[\protect\citeauthoryear{{Mashonkina}, {Gehren}, {Shi}, {Korn} \&
  {Grupp}}{{Mashonkina} et~al.}{2011}]{mash_fe}
{Mashonkina} L.,  {Gehren} T.,  {Shi} J.-R.,  {Korn} A.~J.,    {Grupp} F.,
  2011, \aap, 528, A87

\bibitem[\protect\citeauthoryear{{May}, {Richter} \& {Wichelmann}}{{May}
  et~al.}{1974}]{MRW}
{May} M.,  {Richter} J.,    {Wichelmann} J.,  1974, \aaps, 18, 405

\bibitem[\protect\citeauthoryear{{O'Brian}, {Wickliffe}, {Lawler}, {Whaling} \&
  {Brault}}{{O'Brian} et~al.}{1991}]{BWL}
{O'Brian} T.~R.,  {Wickliffe} M.~E.,  {Lawler} J.~E.,  {Whaling} W.,
  {Brault} J.~W.,  1991, Journal of the Optical Society of America B Optical
  Physics, 8, 1185

\bibitem[\protect\citeauthoryear{{Piskunov} \& {Valenti}}{{Piskunov} \&
  {Valenti}}{2015}]{SME2015ApJ}
{Piskunov} N.,  {Valenti} J.~A.,  2015, \apj, in preparation

\bibitem[\protect\citeauthoryear{{Raassen} \& {Uylings}}{{Raassen} \&
  {Uylings}}{1998}]{RU}
{Raassen} A.~J.~J.,  {Uylings} P.~H.~M.,  1998, \aap, 340, 300

\bibitem[\protect\citeauthoryear{{Ralchenko}, {Kramida}, {Reader} \& {NIST ASD
  Team}}{{Ralchenko} et~al.}{2010}]{NIST10}
{Ralchenko} Y.,  {Kramida} A.,  {Reader} J.,    {NIST ASD Team}, 2010, NIST
  Atomic Spectra Database (ver. 4.0.0), http://www.nist.gov/pml/data/asd.cfm

\bibitem[\protect\citeauthoryear{{Ram{\'{\i}}rez}, {Mel{\'e}ndez} \&
  {Asplund}}{{Ram{\'{\i}}rez} et~al.}{2009}]{2009AA...508L..17R}
{Ram{\'{\i}}rez} I.,  {Mel{\'e}ndez} J.,    {Asplund} M.,  2009, \aap, 508, L17

\bibitem[\protect\citeauthoryear{{Ryabchikova}, {Piskunov}, {Kurucz},
  {Stempels}, {Heiter} \& {et al.,}}{{Ryabchikova}
  et~al.}{2015}]{2015PhyS...90e4005R}
{Ryabchikova} T.,  {Piskunov} N.,  {Kurucz} R.~L.,  {Stempels} H.~C.,  {Heiter}
  U.,    {et al.,} 2015, \physscr, 90, 054005

\bibitem[\protect\citeauthoryear{{Ryabchikova}, {Piskunov}, {Stempels}, {Kupka}
  \& {Weiss}}{{Ryabchikova} et~al.}{1999}]{T83av}
{Ryabchikova} T.~A.,  {Piskunov} N.~E.,  {Stempels} H.~C.,  {Kupka} F.,
  {Weiss} W.~W.,  1999, \physscr Volume T, 83, 162

\bibitem[\protect\citeauthoryear{{Saar} \& {Osten}}{{Saar} \&
  {Osten}}{1997}]{1997MNRAS.284..803S}
{Saar} S.~H.,  {Osten} R.~A.,  1997, \mnras, 284, 803

\bibitem[\protect\citeauthoryear{{Santos}, {Israelian} \& {Mayor}}{{Santos}
  et~al.}{2004}]{2004AA...415.1153S}
{Santos} N.~C.,  {Israelian} G.,    {Mayor} M.,  2004, \aap, 415, 1153

\bibitem[\protect\citeauthoryear{{Shulyak}, {Tsymbal}, {Ryabchikova},
  {St{\"u}tz} \& {Weiss}}{{Shulyak} et~al.}{2004}]{2004AA...428..993S}
{Shulyak} D.,  {Tsymbal} V.,  {Ryabchikova} T.,  {St{\"u}tz} C.,    {Weiss}
  W.~W.,  2004, \aap, 428, 993

\bibitem[\protect\citeauthoryear{{Sitnova}, {Mashonkina} \&
  {Ryabchikova}}{{Sitnova} et~al.}{2015}]{ti_atom}
{Sitnova} T.,  {Mashonkina} L.,    {Ryabchikova} T.,  2015, in preparation

\bibitem[\protect\citeauthoryear{{Sitnova}, {Zhao}, {Mashonkina}, {Chen},
  {Liu}, {Pakhomov}, {Tan}, {Bolte}, {Alexeeva}, {Grupp}, {Shi} \&
  {Zhang}}{{Sitnova} et~al.}{2015}]{2015ApJ...808..148S}
{Sitnova} T.,  {Zhao} G.,  {Mashonkina} L.,  {Chen} Y.,  {Liu} F.,  {Pakhomov}
  Y.,  {Tan} K.,  {Bolte} M.,  {Alexeeva} S.,  {Grupp} F.,  {Shi} J.-R.,
  {Zhang} H.-W.,  2015, \apj, 808, 148

\bibitem[\protect\citeauthoryear{{Smiljanic}, {Korn}, {Bergemann}, {Frasca},
  {Magrini} \& {et al.,}}{{Smiljanic} et~al.}{2014}]{2014A&A...570A.122S}
{Smiljanic} R.,  {Korn} A.~J.,  {Bergemann} M.,  {Frasca} A.,  {Magrini} L.,
  {et al.,} 2014, \aap, 570, A122

\bibitem[\protect\citeauthoryear{{Soubiran}, {Le Campion}, {Cayrel de Strobel}
  \& {Caillo}}{{Soubiran} et~al.}{2010}]{2010AA...515A.111S}
{Soubiran} C.,  {Le Campion} J.-F.,  {Cayrel de Strobel} G.,    {Caillo} A.,
  2010, \aap, 515, A111

\bibitem[\protect\citeauthoryear{{Sousa}, {Santos}, {Israelian}, {Mayor} \&
  {Monteiro}}{{Sousa} et~al.}{2006}]{2006AA...458..873S}
{Sousa} S.~G.,  {Santos} N.~C.,  {Israelian} G.,  {Mayor} M.,    {Monteiro}
  M.~J.~P.~F.~G.,  2006, \aap, 458, 873

\bibitem[\protect\citeauthoryear{{Sousa}, {Santos}, {Mayor}, {Udry},
  {Casagrande} \& {et al.,}}{{Sousa} et~al.}{2008}]{2008AA...487..373S}
{Sousa} S.~G.,  {Santos} N.~C.,  {Mayor} M.,  {Udry} S.,  {Casagrande} L.,
  {et al.,} 2008, \aap, 487, 373

\bibitem[\protect\citeauthoryear{{Tanner}, {Boyajian}, {von Braun}, {Kane},
  {Brewer} \& {et al.,}}{{Tanner} et~al.}{2015}]{2015ApJ...800..115T}
{Tanner} A.,  {Boyajian} T.~S.,  {von Braun} K.,  {Kane} S.,  {Brewer} J.~M.,
   {et al.,} 2015, \apj, 800, 115

\bibitem[\protect\citeauthoryear{{Torres}, {Fischer}, {Sozzetti}, {Buchhave},
  {Winn} \& {et al.,}}{{Torres} et~al.}{2012}]{2012ApJ...757..161T}
{Torres} G.,  {Fischer} D.~A.,  {Sozzetti} A.,  {Buchhave} L.~A.,  {Winn}
  J.~N.,    {et al.,} 2012, \apj, 757, 161

\bibitem[\protect\citeauthoryear{{Torres}, {Winn} \& {Holman}}{{Torres}
  et~al.}{2008}]{2008ApJ...677.1324T}
{Torres} G.,  {Winn} J.~N.,    {Holman} M.~J.,  2008, \apj, 677, 1324

\bibitem[\protect\citeauthoryear{{Tsymbal}}{{Tsymbal}}{1996}]{synthV}
{Tsymbal} V.,  1996, in {Adelman} S.~J.,  {Kupka} F.,   {Weiss} W.~W.,  eds,
  M.A.S.S., Model Atmospheres and Spectrum Synthesis Vol.~108 of Astronomical
  Society of the Pacific Conference Series, {STARSP: A Software System For the
  Analysis of the Spectra of Normal Stars}.
p.~198

\bibitem[\protect\citeauthoryear{{Valenti} \& {Fischer}}{{Valenti} \&
  {Fischer}}{2005}]{2005ApJS..159..141V}
{Valenti} J.~A.,  {Fischer} D.~A.,  2005, \apjs, 159, 141

\bibitem[\protect\citeauthoryear{{Valenti} \& {Piskunov}}{{Valenti} \&
  {Piskunov}}{1996}]{1996AAS..118..595V}
{Valenti} J.~A.,  {Piskunov} N.,  1996, \aaps, 118, 595

\bibitem[\protect\citeauthoryear{{van Belle}, {van Belle}, {Creech-Eakman},
  {Coyne}, {Boden} \& {et al.,}}{{van Belle}
  et~al.}{2008}]{2008ApJS..176..276V}
{van Belle} G.~T.,  {van Belle} G.,  {Creech-Eakman} M.~J.,  {Coyne} J.,
  {Boden} A.~F.,    {et al.,} 2008, \apjs, 176, 276

\bibitem[\protect\citeauthoryear{{von Braun}, {Boyajian}, {van Belle}, {Kane},
  {Jones} \& {et al.,}}{{von Braun} et~al.}{2014}]{2014MNRAS.438.2413V}
{von Braun} K.,  {Boyajian} T.~S.,  {van Belle} G.~T.,  {Kane} S.~R.,  {Jones}
  J.,    {et al.,} 2014, \mnras, 438, 2413

\bibitem[\protect\citeauthoryear{{Wood}, {Lawler}, {Sneden} \& {Cowan}}{{Wood}
  et~al.}{2013}]{WLSC}
{Wood} M.~P.,  {Lawler} J.~E.,  {Sneden} C.,    {Cowan} J.~J.,  2013, \apjs,
  208, 27

\end{thebibliography}

\appendix
\section{}
\begin{table*}
\caption{Laboratory wavelengths of the mask m6. The beginning of each region
marked by bold face.}\label{tab:mask}
\centering
\begin{tabular}{llllll}
\hline
\multicolumn{6}{c}{First and last wavelengths, \AA}\\
\hline
\bf 4485.468--4486.101 & 4551.114--4551.445 & 4865.521--4866.407 &
5193.964--5199.393 & 6110.657--6111.972 & 6524.252--6525.690 \\
4487.957--4490.926 & 4551.560--4551.914 & 4867.344--4869.003 & \bf
5597.997--5598.714 & 6112.400--6113.743 & 6526.200--6526.773 \\
4491.260--4491.449 & 4553.698--4553.938 & 4869.354--4869.523 &
5599.274--5600.566 & 6113.933--6115.053 & 6526.966--6527.376 \\
4491.560--4491.961 & 4554.133--4556.343 & 4869.800--4870.212 &
5602.151--5602.412 & 6116.792--6117.098 & 6527.740--6527.972 \\
4492.173--4493.621 & 4556.652--4557.053 & 4870.687--4871.518 &
5602.644--5603.527 & 6117.968--6120.606 & 6528.219--6528.366 \\
4494.423--4494.813 & 4557.970--4558.887 & 4871.863--4875.116 &
5603.968--5607.261 & 6120.898--6122.144 & 6528.807--6529.318 \\
4495.337--4495.660 & 4559.851--4560.367 & 4875.371--4876.600 &
5607.494--5608.483 & 6122.293--6123.008 & 6529.512--6529.876 \\
4495.861--4496.341 & 4560.619--4562.433 & 4877.489--4877.654 &
5608.797--5609.562 & 6123.927--6126.413 & 6530.147--6530.433 \\
4496.731--4497.791 & 4562.548--4562.789 & 4877.948--4879.939 &
5609.899--5613.477 & 6126.884--6129.366 & 6530.712--6530.929 \\
4498.640--4499.288 & 4563.146--4564.076 & \bf 5100.539--5100.757 &
5613.808--5616.797 & 6129.802--6130.990 & 6531.200--6531.804 \\
4499.422--4499.846 & 4564.260--4564.421 & 5101.011--5101.606 &
5617.789--5621.024 & 6131.454--6135.129 & 6532.052--6532.184 \\
4500.215--4501.377 & 4564.640--4565.974 & 5101.916--5104.571 &
5621.453--5625.453 & 6135.293--6135.518 & 6532.664--6533.020 \\
4502.104--4502.350 & 4566.365--4567.918 & 5104.739--5104.962 &
5627.414--5631.505 & 6135.922--6138.042 & 6533.261--6533.331 \\
4502.507--4502.697 & 4568.517--4569.807 & 5105.254--5105.844 &
5632.125--5637.237 & 6139.136--6139.492 & 6533.586--6534.098 \\
4503.637--4504.342 & 4570.153--4570.580 & 5106.105--5107.813 &
5637.911--5639.934 & 6139.903--6141.676 & 6534.408--6534.493 \\
4504.790--4504.947 & 4570.741--4572.505 & 5108.248--5108.708 &
5640.252--5640.403 & 6141.779--6148.506 & 6534.734--6534.850 \\
4506.683--4506.918 & 4573.048--4573.498 & 5109.000--5113.649 &
5640.638--5644.453 & 6149.089--6149.460 & 6535.284--6535.928 \\
4507.053--4507.300 & 4573.913--4574.352 & 5113.948--5114.385 &
5644.765--5648.889 & 6149.632--6153.129 & 6536.037--6536.324 \\
4508.152--4508.443 & 4574.583--4575.253 & 5115.064--5117.333 &
5649.216--5653.301 & 6153.486--6154.167 & 6536.851--6537.022 \\
4509.172--4509.890 & 4575.692--4575.900 & 5117.508--5117.645 &
5653.743--5653.971 & 6154.298--6156.141 & 6537.565--6537.775 \\
4510.743--4510.912 & 4576.131--4576.593 & 5117.795--5124.717 &
5654.855--5655.639 & 6156.616--6157.002 & 6538.100--6540.306 \\
4511.777--4511.990 & 4577.090--4577.275 & 5124.986--5125.550 &
5655.815--5663.368 & 6157.525--6157.862 & 6540.577--6541.137 \\
4512.158--4512.417 & 4577.379--4579.449 & 5125.762--5126.564 &
5663.911--5666.839 & 6158.055--6158.365 & 6541.455--6542.054 \\
4512.653--4513.114 & 4579.970--4580.676 & 5127.096--5131.100 &
5667.100--5671.172 & 6159.164--6160.080 & 6542.434--6542.862 \\
4514.047--4514.575 & 4580.989--4581.730 & 5131.345--5134.043 &
5672.008--5672.621 & 6160.514--6160.679 & 6543.111--6543.367 \\
4515.047--4515.475 & 4582.726--4584.175 & 5135.362--5136.173 &
5672.897--5673.288 & 6160.845--6162.086 & 6544.153--6544.619 \\
4516.195--4516.398 & 4584.500--4587.285 & 5136.355--5140.651 &
5673.533--5674.016 & 6162.279--6163.706 & 6545.093--6545.545 \\
4517.422--4517.681 & 4587.912--4588.481 & 5141.041--5144.103 &
5677.178--5679.466 & 6163.810--6164.458 & 6545.872--6546.868 \\
4517.906--4518.807 & \bf 4820.119--4820.667 & 5144.500--5146.891 &
5679.835--5680.373 & 6164.865--6165.700 & 6547.015--6547.210 \\
4519.371--4520.374 & 4820.905--4821.397 & 5147.295--5150.244 &
5681.980--5682.872 & 6166.094--6166.405 & 6548.214--6548.456 \\
4522.574--4523.545 & 4823.187--4824.351 & 5150.446--5151.324 &
5684.050--5684.681 & 6166.494--6167.634 & 6548.751--6548.860 \\
4523.850--4523.996 & 4825.815--4826.259 & 5151.678--5153.721 &
5686.421--5687.184 & 6167.993--6168.981 & 6549.211--6550.068 \\
4524.516--4524.787 & 4828.851--4829.496 & 5153.981--5155.918 &
5687.823--5688.425 & 6169.105--6169.499 & 6550.442--6550.800 \\
4525.035--4525.431 & 4831.002--4831.308 & 5156.196--5156.462 &
5689.234--5689.535 & 6169.645--6169.845 & 6550.987--6551.938 \\
4525.736--4527.522 & 4831.536--4831.736 & 5157.457--5159.536 &
5689.689--5690.136 & 6170.114--6171.124 & 6552.109--6552.297 \\
4527.737--4527.862 & 4832.309--4832.882 & 5159.669--5160.659 &
5690.337--5690.668 & 6171.324--6171.746 & 6552.866--6553.645 \\
4528.212--4529.220 & 4833.038--4833.288 & 5160.932--5162.461 &
5691.100--5692.135 & 6171.968--6172.341 & 6554.028--6554.995 \\
4530.397--4531.768 & 4834.897--4835.125 & 5163.274--5165.529 &
5694.435--5695.123 & 6172.819--6173.490 & 6555.206--6555.713 \\
4533.117--4533.445 & 4835.777--4836.395 & 5165.714--5166.025 &
5695.996--5696.499 & 6173.739--6177.818 & 6555.931--6557.032 \\
4533.820--4536.668 & 4836.724--4836.947 & 5166.165--5167.252 &
5697.117--5698.099 & 6179.267--6180.334 & 6557.336--6557.992 \\
4537.316--4537.770 & 4837.549--4838.759 & 5167.424--5168.022 &
5698.254--5698.625 & 6180.612--6181.403 & 6558.687--6560.382 \\
4538.520--4539.009 & 4839.406--4842.063 & 5168.506--5172.582 &
5699.832--5700.622 & 6182.867--6183.374 & 6561.195--6561.500 \\
4539.498--4539.907 & 4842.560--4845.746 & 5172.798--5174.163 &
5700.823--5701.729 & 6183.783--6185.429 & 6564.339--6568.645 \\
4540.601--4541.136 & 4847.890--4848.349 & 5174.781--5175.132 & \bf
6099.997--6100.504 & 6186.527--6188.293 & 6568.927--6570.440 \\
4541.420--4541.602 & 4848.791--4849.284 & 5175.324--5178.893 &
6101.693--6102.667 & 6188.724--6189.266 & 6571.059--6571.867 \\
4542.319--4542.809 & 4851.391--4851.604 & 5179.059--5180.996 &
6102.795--6103.796 & 6189.795--6192.071 & 6572.306--6572.479 \\
4543.355--4544.859 & 4852.412--4853.394 & 5182.160--5183.478 &
6104.155--6104.730 & 6192.356--6193.087 & 6572.643--6574.652 \\
4545.030--4545.498 & 4854.707--4854.999 & 5183.753--5187.122 &
6104.953--6105.231 & 6193.965--6194.153 & 6574.942--6575.225 \\
4545.851--4546.079 & 4855.274--4856.572 & 5187.346--5189.046 &
6105.454--6106.280 & 6194.341--6196.007 & 6575.405--6576.112 \\
4546.364--4547.984 & 4857.240--4857.639 & 5189.296--5189.706 &
6106.917--6107.844 & 6196.258--6199.990 & 6576.505--6576.803 \\
4548.658--4548.875 & 4858.921--4860.125 & 5189.905--5191.664 &
6107.987--6108.380 & \bf 6520.359--6521.702 & 6577.039--6579.082 \\ 
4549.343--4549.926 & 4863.312--4865.285 & 5191.914--5193.636 &
6109.220--6110.176 & 6523.000--6523.217 & 6579.475--6579.781 \\

\hline
\end{tabular}
\end{table*}

\bsp    
\label{lastpage}
\end{document}